%% file: issta.tex
\documentclass[sigconf,screen]{acmart}

\setcopyright{rightsretained}
\acmPrice{}
\acmDOI{10.1145/3597926.3598148}
\acmYear{2023}
\copyrightyear{2023}
\acmSubmissionID{issta23main-p426-p}
\acmISBN{979-8-4007-0221-1/23/07}
\acmConference[ISSTA '23]{Proceedings of the 32nd ACM SIGSOFT International Symposium on Software Testing and Analysis}{July 17--21, 2023}{Seattle, WA, United States}
\acmBooktitle{Proceedings of the 32nd ACM SIGSOFT International Symposium on Software Testing and Analysis (ISSTA '23), July 17--21, 2023, Seattle, WA, United States}
\received{2023-02-16}
\received[accepted]{2023-05-03}



\usepackage{microtype}
\usepackage{bm}
\usepackage[normalem]{ulem}
\usepackage[inline]{enumitem}
\usepackage[english]{babel}
\usepackage{algorithm}
\usepackage{algpseudocode}
\algblockdefx[doWhile]{Do}{DoWhile}{\textbf{do}}[1]{\textbf{while} #1}

\usepackage{listings}
\usepackage{graphics}
\usepackage{tikz}
\usetikzlibrary{positioning}
\usepackage{colortbl}
\usepackage{url}
\usepackage{pifont}
\usepackage{cleveref}
\newcommand*\mean[1]{\overline{#1}}
\newcommand{\p}{\ding{51}}
\newcommand{\f}{\ding{55}}
\usepackage{changepage}

\newcommand{\commentedout}[1]{\{\ldots\}}
\newcommand{\outdated}[1]{}
\newcommand{\cupdot}{\mathbin{\mathaccent\cdot\cup}}
\newcommand{\tp}{\cellcolor{yellow!50}}
 \newcommand{\mysubsubsection}[1]{\smallskip\noindent\emph{#1}.}
\definecolor{grey}{rgb}{0.95,0.95,0.95}
\definecolor{boxgreen}{rgb}{0.89,0.94,0.85}
\definecolor{greyborder}{rgb}{0.78,0.78,0.78}
\definecolor{codegreen}{rgb}{0,0.6,0}
\definecolor{codegray}{rgb}{0.5,0.5,0.5}
\definecolor{codepurple}{rgb}{0.58,0,0.82}
\definecolor{backcolour}{rgb}{1.0,1.0,0.7}

\lstdefinestyle{mystyle}{%
  commentstyle=\color{codegreen},
  keywordstyle=\color{magenta},
  numberstyle=\tiny\color{codegray},
  stringstyle=\color{codepurple},
  basicstyle=\ttfamily\scriptsize\linespread{0.85}\selectfont,
  breakatwhitespace=false,
  breaklines=true,
  captionpos=b,
  keepspaces=true,
  numbers=left,
  numbersep=5pt,
  showspaces=false,
  showstringspaces=false,
  showtabs=false,
  tabsize=2
}
\newcommand*\Let[2]{\State #1 $\gets$ #2}

\begin{document}
\title{Improving Spectrum-Based Localization of Multiple Faults by Iterative Test Suite Reduction}
\author{Dylan Callaghan}
\affiliation{%
  \institution{Stellenbosch University}
  \city{Stellenbosch}
  \country{South Africa}
}
\email{21831599@sun.ac.za}
\author{Bernd Fischer}
\affiliation{%
  \institution{Stellenbosch University}
  \city{Stellenbosch}
  \country{South Africa}
}
\email{bfischer@sun.ac.za}

\begin{abstract}
%
Spectrum-based fault localization (SBFL) works well for single-fault programs but its accuracy decays for
increasing fault numbers.
We present
FLITSR (Fault Localization by Iterative Test Suite Reduction),
a novel SBFL extension
that improves the
localization of
a given base metric specifically in the presence
of multiple faults.  FLITSR iteratively selects reduced versions of the test
suite that better localize the individual faults in the system. This allows
it to identify and re-rank faults
ranked too low by
the base metric because they were masked by other program elements.


We evaluated FLITSR over method-level spectra from an existing large synthetic
dataset comprising 75000 variants of 15 open-source projects with up to 32
injected faults, as well as method- and statement-level spectra from a new
dataset with 326 true multi-fault versions from the Defects4J benchmark set
containing up to 14 real faults.  For all three spectrum types we consistently
see substantial reductions of the average wasted efforts at different fault
levels, of 30\%-90\% over the best base metric, and generally similarly large
increases in precision and recall, albeit with larger variance across the
underlying projects.  For the method-level real faults, FLITSR also
substantially outperforms GRACE, a state-of-the-art learning-based fault
localizer.

\end{abstract}
\maketitle
\input{intro}
\input{background}

\input{approach}
\input{mult}

\input{eval}
\input{relwork}

\input{conclude}
\begin{acks}
D.~Callaghan was supported by impact.com, the Skye foundation and the SU
Postgraduate Scholarship Programme (PSP).
\end{acks}
\newpage
\bibliographystyle{ACM-Reference-Format}
\bibliography{refs}
\end{document}

%% file: intro.tex
\section{Introduction}
\label{sec:intro}
%
Debugging (i.e., finding and fixing bugs) contributes to a large part of the
costs of the software development process~\cite{VESSEY1985459}, and substantial
efforts have been made to automate it
\cite{survey,flfamilies}.
Debugging involves
fault localization, fault understanding, and
fault correction~\cite{DBLP:conf/issta/ParninO11}; we focus on the first here.
There are many approaches to automated fault localization, most notably
learning-based~\cite{grace, deepFl, DBLP:journals/tr/WongDGXT12},
model-based~\cite{survey,modelSurvey}, and
\emph{spectrum-based fault localization} (SBFL)~\cite{sbflSurvey, tarantula, ochiai,
dstar}. SBFL is a lightweight and scalable approach that
%
summarizes coverage information for each program element over a given test
suite into a spectrum, and uses a given formula or \emph{metric} to compute
from the spectrum a \emph{suspiciousness score}, which is seen as likelihood
for elements to contain a fault.



SBFL has been widely studied, with a variety of different metrics and extensions
suggested (see~\Cref{sec:relwork}), and with successful applications
to downstream tasks such as automated program repair~\cite{repairDebroyWong,
repairClaire, repairSemfix}.
In several studies involving single-fault programs, users of SBFL tools were found
to be more effective (i.e.,  found more bugs)~\cite{gzoltarUserStudy, sbflUserStudy,
davidUserStudy, application} and more efficient (i.e., found bugs faster)~\cite{gzoltarUserStudy,
sbflUserStudy, interactiveUserStudy} than users without such tools.
SBFL has been shown to perform well when there are only a few test failures related to a
single fault~\cite{eval}.
However, its effectiveness degrades,
in particular beyond the first fault, when there are
multiple test failures related to multiple faults
\cite{multipleInfluence}. This
\emph{multi-fault localization} (MFL) problem naturally emerges in many situations,
e.g., after testing by external QA, fuzzing campaigns, or field releases of
new versions, or simply through the accumulation of unresolved issues.

Several approaches have been proposed
to address
the MFL problem
\cite{DBLP:journals/infsof/ZakariLAAR20}.
The most widely used one is the \emph{one-bug-at-a-time} (OBA) approach
\cite{tarantula,DBLP:conf/issre/DebroyW09,dstar,DBLP:journals/tsmc/WongDX12,DBLP:conf/kbse/JonesH05},
which
iteratively finds and fixes the top-ranked faults.
%
However, OBA is not realistic in practice, where bugs are fixed based on priorities
other than their SBFL rank (e.g., system impact, time to fix, availability
of developer responsible for the code, etc.).
Alternatively, in the \emph{parallel debugging} approach the spectrum is decomposed into
several blocks or \emph{fault-focused clusters} where failures caused
by the same fault are grouped into the same cluster
\cite{parallel:jones,DBLP:journals/tr/WongDGXT12,DBLP:journals/tse/GaoW19,ILP}.
However, due to fault interference and fault masking the decompositions are
often imperfect~\cite{parallel:steimann}, so that clusters again induce an MFL problem.
%
True \emph{multiple-bug-at-a-time} (MBA) approaches
localize multiple faults in one run, without any changes to the target
system.
%
However, MBA has ``enjoyed little attention''
\cite{DBLP:journals/infsof/ZakariLAAR20}, with only few approaches published
\cite{DBLP:journals/jss/AbreuZG11, zoltarM, DBLP:journals/access/ZakariLC18,
DBLP:journals/jss/ZhengWFCY18}. Most of these require exceedingly large runtimes
due to their incorporation of various other technologies such as SAT
solvers~\cite{zoltarM}.

In this paper, we address this gap and present a novel, purely spectrum-based
MBA approach called FLITSR (\emph{Fault Localization by Iterative Test Suite
Reduction}).
It is based on the observation that in a ranking induced by an SBFL metric, the
position of faulty elements is influenced by the density of failing tests
executing each of these faults. Our key insight is that by removing from the test
suite the failing tests executing the most suspicious element, we can increase the
relative density of failing tests executing the remaining faulty elements, and so
improve the ranks of these faults. We can thus use any SBFL metric in a feedback
loop (alternating between localizing and recording the most suspicious element and
removing its associated failing tests) in order to reduce the test
suite in each iteration to localize the remaining faults more accurately.
%
%
Through this process, FLITSR constructs a set of highly suspicious program
elements called a \emph{basis},
where the execution of each failing test in the test suite involves at least one
basis element. The basis element(s) executed in this test can thus be seen as the
cause of the test failure, and the constructed basis thus
fully \emph{explains} all test failures.
FLITSR then computes a new ranking by shifting the basis elements
towards the top, since it considers them as the elements most likely to contain
a fault.
For most program spectra, multiple bases exist; we exploit this in a
variant of FLITSR called FLITSR*, which iteratively calculates the various
bases, and ranks them in the order that they were found.
%
This sequence of bases also gives us a sequence of natural cut-off points (i.e., the end of each
basis in the ranking), where developers can choose to stop inspecting the
ranking as further inspection yields diminishing returns. This is in
line with how developers use SBFL tools~\cite{DBLP:conf/issta/ParninO11}.

\outdated{
that the ranking induced by a given SBFL metric is
influenced by the density of failing tests executing the individual faults in a
multi-fault system. Our key insight is that we can exploit this property using
the original ranking to iteratively reduce the test suite for each fault, so
that we can more precisely identify multiple faults.
More specifically, FLITSR uses these reduced test suites to construct a set of
faulty program elements called a \emph{basis} (see~\Cref{sec:background}) that
fully explains all test failures (i.e., all failures can be attributed to one
or more elements in the basis).
FLITSR then computes a new ranking by shifting the basis elements towards the
top, since it has deemed them the elements most likely to be faulty.
This ranked basis gives us natural cut-off points (i.e., the end of each
basis in the ranking), where developers can choose to stop inspecting the
ranking as further inspection will produce only diminishing returns. This is in
line with how developers use SBFL tools~\cite{DBLP:conf/issta/ParninO11}.
For most program spectra, multiple bases exist; we exploit that structure in a
variant of FLITSR called FLITSR*, which iteratively calculates the various bases,
and ranks them in the order that they were found by FLITSR.
}

\outdated{
===========

[It is not the same as debugging in parallel, which is based on spectrum decomposition]
===========

[Other basis-based approaches]

===========
}

FLITSR is generic in the SBFL metric used and extends any metric's induced localization
to the MFL scenario, even if the given metric was originally developed for
\emph{single} faults.
%
%
We experimentally evaluate FLITSR using eleven
well-performing
SBFL metrics including popular single-fault metrics~\cite{tarantula, ochiai,
dstar, gp13, wong, naish2},
state-of-the-art multi-fault metrics~\cite{parallel:steimann,zoltarM,barinel,hyperbolic},
and learning-based metrics~\cite{grace}, and show that it unanimously improves on
all base metrics, providing the best overall ranking using metrics that perform best
in the single-fault case. This shows that FLITSR indeed
successfully extends single-fault solutions to the multi-fault case.

The current evaluation of MFL techniques poses significant problems, due to the
scarcity of suitable datasets with known fault locations.
We first use an existing set of \emph{test coverage matrices} (TCMs)
\cite{parallel:steimann, tcm, threats} that has also been used in other projects
\cite{threats,ILP,DBLP:conf/icst/SunP16},
with up to $N_f=32$ injected faults.
Here, the best FLITSR*-variant
reduces the average wasted effort to the first resp.\ median fault
by 20\% resp.\ 40\% ($N_f=4$) to
35\% resp.\ 45\% ($N_f=32$), compared to the best base metric,
and increases precision by 60\%, and recall by 40\% to 75\%.

We also contribute a new multi-fault dataset derived from the Defects4J dataset
\cite{defects4j} of real bugs, and use it to evaluate FLITSR at the method and statement
levels.
This is the first large multi-fault dataset that
contains only existing real faults.%
\footnote{	Note that the current literature uses the default dataset that is
overwhelmingly comprised of single-fault versions, and contains only some
accidental multi-fault versions (e.g.,
parallel patches in cloned methods).
}
For this dataset, the results vary considerably between the individual
projects, and in fact more than between method- and statement-level.
For two projects (Chart and Lang),
FLITSR yields substantial (30\%-90\%)
method-level wasted effort reductions,
and similar increases in
recall (30\%) and precision (20\%-50\%),
and can pinpoint the top error in about half of the cases.
For the remaining three projects, FLITSR* can improve on FLITSR,
with improvements over the base metric for all measures in almost all cases.
%
At the statement-level, the overall results are similar, but the improvements are
relatively smaller, due to the larger sizes of the spectra.
We finally show that FLITSR outperforms by a wide margin
GRACE \cite{grace}, a state-of-the-art
learning-based method-level fault localizer.

In
this
paper we make the following original contributions:
\begin{itemize}
  \item We present a novel SBFL technique called FLITSR, and its
    extension, FLITSR*, that localize \emph{multiple bugs at a time}.
  \item We evaluate our approach compared to eleven popular and well-performing
    SBFL metrics, including three that have been designed for multi-fault
    localization. We demonstrate large improvements for multi-fault problems.
  \item We show that FLITSR outperforms GRACE \cite{grace}, a state-of-the-art
    learning-based fault localizer.
  \item We provide a replication package, with a complete working implementation
    of our approach, and a new real multi-fault version of the Defects4J dataset.
\end{itemize}

%% file: background.tex
\section{Spectrum-Based Fault Localization}%
%
\label{sec:background}
In this section we give background on SBFL and formalize its basic
notations; we also introduce the concepts of \emph{fault-covering span} and
\emph{basis}, which are at the core of FLITSR.

\mysubsubsection{System under test and test suites}
A \emph{system under test} (SUT) $S=\{e_1,\ldots,e_m\}$ consists
of \emph{program elements} $e_i$ at a given level of \emph{coverage
granularity} (typically statements, basic blocks, methods, classes, or
packages).
For SBFL, we require as input the SUT $S$, a \emph{test suite} $T=\{t_1,\ldots,t_n\}$,
%
and an (execution) \emph{oracle} that labels the execution $S(t)$ of the
SUT over each test $t$ as either \emph{failing} or \emph{passing} and so partitions $T$
into failing and passing tests $F$ and $P$, respectively.
We identify the coverage spectra for the execution of the SUT over a test with the
set of executed program elements
(i.e., $S(t)\subset S$), and define the set of failing (resp.\ passing) tests
of an element $e\in S$ as $F_{S,T}(e)=\{t\in F\mid e\in S(t)\}$ (resp.\ $P_{S,T}(e)=\{t\in
P\mid e\in S(t)\}$). We drop the subscripts on $F_{S,T}$ resp.\ $P_{S,T}$ if $S$ resp.\ $T$ are clear from the
context, and define $F(E)=\bigcup_{e\in E}F(e)$
as the failing tests of a set of elements $E$.
For the evaluation, we require a \emph{localization oracle} that labels a subset
$B\subset S$ of program elements (``bugs'') as \emph{faulty} and so provides ground
truth.

\mysubsubsection{Spectrum-based fault localization}
SBFL uses coverage information for the SUT's elements to localize faults, which is
collected when the SUT is executed over the test suite. SBFL techniques typically
do not work directly with the full \emph{execution spectrum} (i.e., the matrix shown
in the upper part of \Cref{tab:motv1}) but with more compact \emph{execution counts}
that count in how many passing (resp.\ failing) test cases an element $e$ is executed
(resp.\ is not executed).
%
SBFL techniques then use a formula called a (ranking) \emph{metric} $M_{S,T}$
to compute from the execution counts a \emph{suspiciousness score} for each
program element. Elements with a higher score are seen as more likely to be
faulty.
A description of the counts as well as a comprehensive list of metrics is
given by Heiden et~al.~\cite{eval}.
We additionally require that the SBFL metric assigns a higher score to
elements that are executed in failing tests than to elements that are not
executed in failing tests. i.e., $F(e_1) \neq \emptyset$ and $F(e_2) = \emptyset$
imply $M_{S,T}(e_1) > M_{S,T}(e_2)$.

%

\mysubsubsection{Ranking and ties}
The elements can be sorted by descending score $M_{S,T}$ order to produce a
suspiciousness \emph{ranking}.
We use $M^\#_{S,T}(e)$ to denote the (ordinal) rank of an element $e$ induced by
the ranking metric $M_{S,T}$.
Note that the rank of an element $e$ can improve (i.e., decrease) even if its
score deteriorates (i.e., decreases) for a different system variant $S'$ (e.g.,
after a bug fix) or test suite $T'$, i.e.,
$M_{S',T}(e)<M_{S,T}(e)$
does \emph{not} imply
$M^\#_{S',T}(e)>M^\#_{S,T}(e)$,
and similarly for $T'$.
Elements with the same score form a \emph{tie group} or \emph{tie}, and can only
be arranged in a random order. A tie that contains both faulty and non-faulty elements
is called \emph{critical}. Elements that have identical execution spectra over the
test suite are always tied and form an \emph{ambiguity group}; SBFL metrics cannot
tell the elements in ambiguity groups apart.

\mysubsubsection{Fault exposure and masking}
We call a fault $e\in B$ \emph{exposed} by a test suite $T$ if it is
executed in at least one failing test (i.e., $F(e)\neq\emptyset$), and require
that all faults $B$ identified by the localization oracle are also exposed
by the test suite $T$, i.e., that $T$ is \emph{strong}.
We define \emph{fault masking} as the interaction between two or more faults
that causes the localization of one or more of these faults to become harder
\cite{faultInterference}.  More formally, fault masking occurs when the rank of one
particular fault $b_i$ increases due to the presence of one or more other
faults $b_j \neq b_i$, i.e., if $S'$ is a variant of the system $S$ in which some of the faults
$b_j$ have been fixed and $M^\#_{S,T}(b_i) > M^\#_{S',T}(b_i)$.  Fault
masking makes fault localization harder.
%
\outdated{
We call a system variant $S'=S[e_1\mapsto e_1']$, where an element $e_1$ is
replaced by $e_1'$, a
\emph{repair} for $e_1$ if $e_1\in B$ is a fault that is
exposed by $T$ and $F_{S',T}(e_1')\subsetneq F_{S,T}(e_1)$, i.e.,
the repair is executed in fewer failing tests than the fault.
We say
that the element $e_0\neq e_1$ \emph{masks} the fault $e_1$ if
$F_{S',T}(e_0)\subsetneq F_{S,T}(e_0)$, i.e., if the repair of $e_1$ also
reduces the set of failing test cases executing $e_0$.  Note that a program
element can mask more than one fault.
Fault masking makes fault localization harder.
}
\outdated{%
In the running example, we have a weak oracle, i.e., $B=\{l_6,l_9,l_{23}\}$ and
$T$ is strong enough for $B$, but there are other faults, e.g., $l_2$ and
$l_{22}$, which $T$ does not expose.
The passing and failing tests of, for example the faulty statement $l_6$ are
$P(l_{6})=\{i_{24},i_{26}\}$ and $F(l_{6})=\{c_1,c_4\}$, respectively.
}%

%

\mysubsubsection{Dominators and domination}
We call an element a \emph{dominator} for a set of elements, if 
every test case executes the set of
elements if and only if it executes the dominator.
More formally, given an element $d \in
S$, and a set of elements $E \subset S$ where $d \notin E$, we call $d$ a dominator
for the elements in the set $E$, if for all $e \in E$, $P(e) \cup F(e) \subseteq
P(d) \cup F(d)$.
We also say that $d$ dominates $E$.

\mysubsubsection{Span and basis}
FLITSR counters fault masking by constructing
(in the first phase of the algorithm)
a fault-covering set or \emph{span} of program elements that together are
executed in all failing tests. More precisely, $C\subseteq S$ is a span if
$F(C)=F(S)=F$.
We call $I\subseteq S$ (fault-) \emph{irreducible} if removing an element
$e$ from $I$ also removes at least one associated failing test, i.e.,
if $F(I\setminus\{e\})\subsetneq F(I)$ for all $e\in I$.
%
A span is called a
\emph{basis} if it is fault-irreducible. Hence,
a span $C$ is a basis iff for all $e\in C$,
$C\setminus\{e\}$ is not a span. In the second phase, FLITSR extracts a
basis from the span constructed in the first phase.

\mysubsubsection{Wasted effort}
A ranking can be used to estimate the effort
developers waste on non-faulty elements if they inspect the program
elements in rank order until they find one ore more faults.
%
The \emph{wasted effort} for a given faulty element in the system is defined as
the number of non-faulty elements ranked above
than the faulty element; we do not include faulty elements here
because we do not consider the corresponding inspection effort as wasted.
In general, the wasted effort is a random variable because the faulty element
can be in a critical tie, where the number of non-faulty elements in that tie that
are ranked above is random.  In the single-fault case the expected value of
this random variable or \emph{average wasted effort} (AWE)
is $(n-1)/2$ for an $n$-way tie.
In the multi-fault case with $k$ faulty elements in the
tie, this generalizes to $(n-k)/(k+1)$ for any of the $k$ faults, and thus
$k\cdot(n-k)/(k+1)$ for all faults.
We use $\mathit{AWE}_k$ to denote the average wasted effort of the $k$-th
fault the ranking.  $\mathit{AWE}_1+1$ is also called the \emph{mean first
rank}.  We also define the average wasted effort to the last ($\mathit{AWE}_L= \mathit{AWE}_{N_f}$)
and median ($\mathit{AWE}_M=\mathit{AWE}_{N_f/2}$) fault in the ranking,
where $N_f=|B|$ is the total
number of faults given by the localization oracle.

\mysubsubsection{Precision and recall}
%
Developers typically inspect only a small number of program
elements before giving up with the ranking~\cite{DBLP:conf/issta/ParninO11}. We
therefore calculate {precision} and {recall}
at a fixed
absolute number $X$ of elements in the ranking.
Hence, the \emph{precision at $X$}
($P@X$) is the fraction of the $X$ elements inspected
that are actually faults, i.e., $k/X$ where $k$ is the expected number of faults found
before the cut-off point $X$. Likewise, the \emph{recall at $X$} ($R@X$) is
the fraction of
all faults found at the cut-off $X$, i.e., $k/N_f$. $R@X$ is the relative version of hit@$X$
\cite{DBLP:conf/issta/ParninO11,eval}; it is more appropriate when the number
of faults varies widely across the different subjects.

%% file: approach.tex
\section{FLITSR Core Algorithm}
%
\label{sec:flitsr-core}
In this section we illustrate and describe in detail how FLITSR's core algorithm works.
%
It takes as input a fixed SUT $S$ and initial test suite $T=F\cupdot P$, and a base
metric $M_{S,T}$.
It returns as output a ranked basis $C\subseteq S$ of highly suspicious elements
$e\in S$ that together are executed in all failing tests $F$, i.e., $F(C)=F$,
along with a ranking for these elements.

\input{motiv}

\begin{algorithm}
  \caption{FLITSR core algorithm}\label{alg:flitsr}
{\small
  \begin{algorithmic}[1]
    \Require{$M$}                     \Comment{base SBFL metric}
    \Require{$S$}                     \Comment{system under test}
    \Require{$T=F\cupdot P$}          \Comment{initial test suite}
    \Ensure{$C$}                      \Comment{span / basis}
    \Ensure{$R$}                      \Comment{FLITSR ranking}
    \Let{$C$}{$\emptyset$}
    \Let{$A$}{$\emptyset$}  \Comment{phase 2 test accumulator}
    \State{{\sc flitsr}($T$)}   \Comment{start phase 1}
    \State{{\sc compact}($R$)}  \Comment{remove gaps in ranking}%
  \vspace*{0.25ex}
  \Function{flitsr}{$T_i$}
    \Comment{{\small\textbf{pre~} $F_{T_i}\!\neq\!\emptyset \wedge F=F_{T_i}\cup F(C)$}}
    \Let{$r$}{$M^\#_{S, T_i}$}         \Comment{rank $S$ over $T_i$ according to $M$}
    \Let{$T_{i+1}$}{\mbox{\sc reduce}($T_i$, $r$)}
    \Comment{{\small\textbf{inv~} $F_{T_{i+1}} \subsetneq F_{T_i} \wedge F=F_{T_{i+1}}\cup F(C)$}}
    \If{$F_{T_{i+1}}\neq \emptyset$}
  \State{{\sc flitsr}($T_{i+1}$)}
    \EndIf                            \Comment{{\small\textbf{inv~}$F=F(C)$}}
    \State{{\sc sift}($T_i$, $r$)}\Comment{continue with phase 2}
    \EndFunction                      \Comment{{\small\textbf{post~}$F=F(C)$}}%

  \vspace*{0.25ex}
  \Function{reduce}{$T_i$, $r$}
    \Let{$e$}{$r$[1]}                 \Comment{select top-ranked element}
    \Let{$C$}{$C \cup \{e\}$}
    \State\Return{$T_i \setminus F_{S, T_i}(e)$}
  \EndFunction%
  \vspace*{0.25ex}
  \Function{sift}{$T_i$, $r$}        \Comment{{\small\textbf{inv~}$F=A \cup F(C)$}}
    \Let{$e$}{$r$[1]}                 \Comment{select top-ranked element}
    \If{$F_{S,T_i}(e) \subseteq A$}
      \Let{$C$}{$C \setminus \{e\}$}
    \Else
    \Let{$R[e]$}{$i$}		      \Comment{set FLITSR-rank of element}
    \Let{$A$}{$A \cup F_{S,T}(e)$}    \Comment{Add failing tests from full test suite}
    \EndIf
  \EndFunction
  \end{algorithmic}
}
\end{algorithm}

\subsection{Phase I: Span Computation}
The FLITSR core algorithm (see \Cref{alg:flitsr}) works in two phases.
In the first phase, it constructs a fault-covering element set or \emph{span},
alternating over two steps until there are no further failing tests:
\begin{description}[topsep=3pt, partopsep=0pt]
  \item[\hspace*{1em}\normalfont\emph{localize}:] use the given base metric
     with the current test suite to identify the most suspicious program element;
     and
  \item[\hspace*{1em}\normalfont\emph{reduce}:] remove from the current test suite
     all failing tests that execute this program element.
\end{description}
The recursive top-level function {\sc flitsr} (lines 5--12) implements
this first phase. It takes as input a current test suite $T_i$ and uses this to rank
the elements in the SUT $S$ (line 6) according to the metric $M$.
It then calls the function {\sc reduce} (lines 13--17), which selects the
most suspicious element $e$, adds it to the span $C$, and removes all
failing tests $F_{S, T_i}(e)$ that execute $e$. Note that $e$ can
be in a tie; if this forms an ambiguity group over the \emph{full} test suite, we
select all elements; otherwise, we use a \emph{tie breaker}
strategy
that gives preference to the original ranking and test suite, and has worked well in practice.%
\footnote{
  Specifically,
  (1) we first select the element with the highest original SBFL score,
    computed over the original test suite;
  (2) if the tie remains, we select the element with the highest number of
    associated failing test cases in the original test suite;
  (3) if the tie still remains, we select the element that occurs lexically
    earliest in the SUT.
}
If the test suite returned by {\sc reduce} still contains failing tests
(line 8), {\sc flitsr} recurses (line 9), otherwise $C$ is already a span, and
{\sc flitsr} proceeds to the {\sc sift} function to start the second
phase (line 11).
Note that FLITSR reduces the test suite only for localization purposes and never re-executes it.

\outdated{
Note that we can deduce from the spectrum that the most suspicious element
(i.e., $l_{12}$) cannot be the single fault in the system because it is
executed only in three of the eight failing tests. We can construct a
fault-covering set or \emph{span} of program elements that together are
executed in all failing tests, e.g., by adding $l_{5}$ and $l_{19}$. However,
these have relatively low suspiciousness scores, and the failing tests that
execute $l_{5}$ also execute $l_{12}$, which introduces redundancy.
In its first phase, FLITSR therefore picks the most suspicious program element,
removes all failing tests that execute this, and then uses the reduced test
suite for the next iteration.  This produces small, redundancy-free spans of
highly suspicious elements.
}

\begin{figure}[t]
  \includegraphics[scale=0.50]{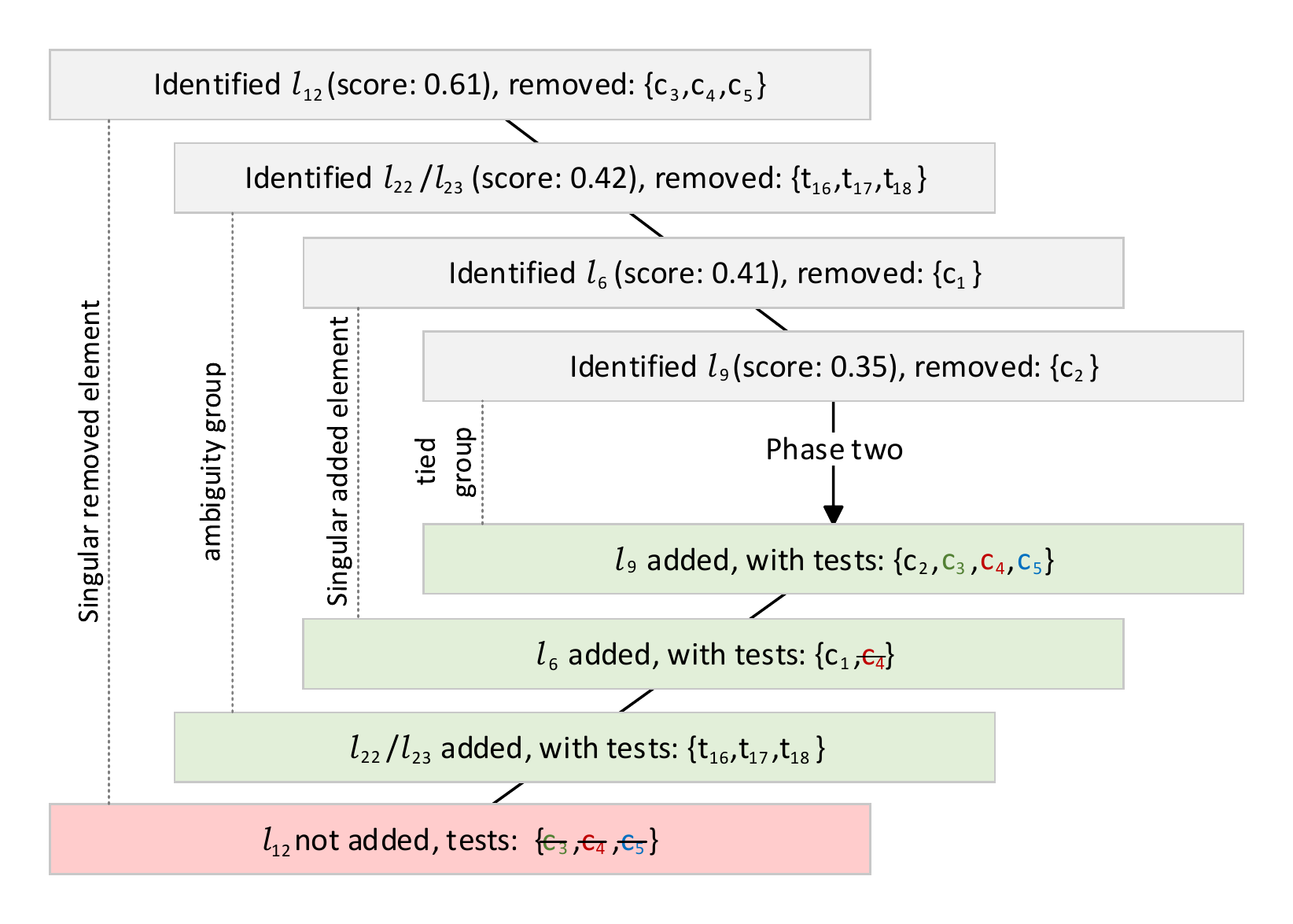}
  \caption{Recursion chain for FLITSR with Ochiai as base metric, over the
    test suite in~\Cref{tab:motv1}.
  }
  \label{fig:exe1}
\end{figure}

\mysubsubsection{Running example}
%
The upper half of \Cref{fig:exe1} illustrates the span computation for the running
example,
using Ochiai as base
metric; \Cref{tab:exeoch} shows the scores in more detail.

\begin{figure}
\setlength{\tabcolsep}{1pt}
\resizebox{\columnwidth}{!}{%
  \begin{tabular}{|c|c|c|c|c|c|c|c|c|c|c|c|c|c|c|c|c|c|c|c|}\hline
    Iteration
     & $l_2$ & $l_3$ & $l_4$ & $l_5$ & \textbf{\emph{l}}$_\mathbf{6}$ & $l_7$
     & $l_8$ & \textbf{\emph{l}}$_\mathbf{9}$ & $l_{10}$ & $l_{12}$
     & $l_{15}$ & $l_{19}$ & $l_{20}$ & $l_{22}$
     & \textbf{\emph{l}}$_\mathbf{23}$ & $l_{24}$ & $l_{25}$ & $l_{26}$ & $l_{28}$\\
    \hline
    1&0.42&0.42&0.42&0.42&0.35&0.37&0.13&0.43&0.13&\tp0.61&0.18&0.32&0.00&0.34&0.34&0.18&0.16&0.00&0.18\\
    2&0.23&0.23&0.23&0.23&0.26&0.13&0.00&0.16&0.16&-&0.23&0.40&0.00&\tp\textbf{0.42}&\tp\textbf{0.42}&0.22&0.20&0.00&0.22\\
    3&0.37&0.37&0.37&0.37&\tp\textbf{0.41}&0.20&0.00&0.25&0.25&-&0.37&0.00&0.00&-&-&0.00&0.00&0.00&0.00\\
    4&0.27&0.27&0.27&0.27&-&0.29&0.00&\tp\textbf{0.35}&\tp0.35&-&0.27&0.00&0.00&-&-&0.00&0.00&0.00&0.00\\
    \hline
    FLITSR &-&-&-&-&\tp\#2&-&-&\tp\#3&-&-&-&-&-&\tp\#1&\tp\#1&-&-&-&-\\\hline
  \end{tabular}
}
\caption{Ochiai scores for each FLITSR iteration in running example;
  the highest scores in each iteration are highlighted, and elements in the
  final basis boldfaced. Final ranking shown at bottom.
}
\label{tab:exeoch}
%
\end{figure}

In the first iteration, Ochiai identifies $l_{12}$ as most suspicious; it is
executed in the failing tests $c_{3}$, $c_{4}$, and $c_{5}$, which are thus removed from the test
suite. In the second iteration, $l_{22}$ and $l_{23}$ both emerge as most suspicious
over this reduced test suite.
%
Note that $l_{22}$ and $l_{23}$ have identical execution spectra over the
original test suite (see \Cref{tab:motv1}) and form an ambiguity group.
Hence, FLITSR adds both to the span, and removes 
$t_{16}$--$t_{18}$.
The third iteration identifies $l_{6}$ and removes $c_{1}$.
The fourth iteration identifies $l_{9}$ and $l_{10}$ as most suspicious;
however, these do \emph{not} form an ambiguity group over the
original test suite and FLITSR applies its tie breaker rule, to pick $l_{9}$
only, since its original Ochiai score is higher than that of $l_{10}$ (see
\Cref{tab:motv1}). After $c_{2}$ is removed, the test suite contains no further
failing tests, and the first phase is completed.

\subsection{Phase II: Span Reduction}

In the second phase, FLITSR \emph{minimizes} the fault-covering set by
eliminating redundant program elements, i.e., by identifying elements which can
be removed from the set without compromising its fault-covering property.
Elements identified in later iterations are more often required in the basis
since they explain failing test cases that elements
from previous iterations did not. FLITSR therefore
%
backtracks over the iterations from the first phase and
checks in
{\sc sift} (lines 18--26) whether
the elements added to the span in the current iteration are subsumed (i.e.,
their failing tests are already accumulated) by elements added in later iterations
(line 20) and can thus be sifted out (line 21).
However, if this is \emph{not} the case, FLITSR considers $e$ to be faulty,
and {\sc sift} sets the rank of $e$ and
adds its failing tests from the \emph{original} test suite to $A$, so that it
accumulates all failing tests of the elements
kept in the basis (line 24).  This ensures that
$C$ maintains the spanning property.

%

The basis elements are then placed at the top of the SBFL ranking, and are
given an internal ranking in the order that they were selected by FLITSR. This
internal ranking is based on the intuition that elements selected earlier by FLITSR
are chosen based on a larger part of the test suite, and are therefore more likely
to be a correct choice.  The internal basis ranking also reduces the number of
large ties in the ranking, which is a difficult problem in
SBFL~\cite{ties1}.

\mysubsubsection{Running example}
%
The lower half of \Cref{fig:exe1} illustrates the span reduction for the running
example.
It starts with the results of the fourth-round iteration of the first phase, i.e.,
the element $l_{9}$, and the failing tests $c_2$--$c_5$; note that the second
phase always uses
the failing tests from the original test suite, not the reduced versions.
Since the accumulator $A$ is initially empty, we keep
$l_{9}$ at rank \#4 and add $c_2$--$c_5$ to $A$.
From the third iteration
we keep $l_{6}$ in the basis at rank \#3, and add $c_{1}$ to $A$ (since it already contains $c_{4}$);
from the second iteration, we keep $l_{22}$ and
$l_{23}$ (since we cannot resolve the ambiguity group) at rank \#2, and add the
tests $t_{16}$--$t_{18}$ to $A$.
Hence, when we return to the first iteration, $A$ already contains all
failing tests that execute $l_{12}$, which means that it is reducible and can be
removed. The second phase therefore ends
with only the four suspicious elements $l_{6}$, $l_{9}$, $l_{22}$, and $l_{23}$.


The final call to {\sc compact} (line 4) simply removes the gaps left in
$R$ by elements not kept in the basis, and ensures the dense ranking shown in
the final line of \Cref{tab:exeoch}. This gives an AWE for the last fault of
0.5, an improvement over all base metrics (see
\Cref{tab:motv1}).

\subsection{Correctness}

It is easy to see that the algorithm terminates in linear time. The SBFL metric
ensures that the top-ranked element $e$ is failing at least one test (see
\Cref{sec:background}) i.e., $F_{S,T_i}(e)\neq\emptyset$, unless all tests remaining
in $T_i$ pass (i.e., $T_i\cap F=\emptyset$). Thus {\sc reduce} removes a non-empty
subset from $T_i$ on every call (i.e., the first part of the invariant on line~7
holds), as long as it contains at least one failing test. Since $T$ is finite, the
condition on line~8 will become true in at most $T$ steps, at which point the
recursion terminates, and the second phase will execute at most $T$ steps as
well.

%

It is also easy to see that the algorithm computes a fault-covering span.
{\sc reduce} maintains the second part of
the invariant on line~7 because it adds $e$ to $C$, so that the tests removed
from $T_i$ are captured in $T_{i+1}$ by $F(C)$. When the condition on line~8
fails, this simplifies to the invariant at line~10.
{\sc sift} maintains a similar invariant (i.e., $F=A\cup F(C)$,
see line~18) in a similar fashion.
Note that {\sc reduce} must remove all remaining failing tests
$F_{S,T_i}(e)$ to ensure correctness;
if it removes a subset $F'\subsetneq F_{S,T_i}(e)$, {\sc sift} is not guaranteed to compute a basis;
if it removes a superset $F_{S,T_i}(e)\subsetneq F'\subset F$, $C$ is not guaranteed to be a span.
%
%
%
%
We can finally show that the span returned by Algorithm~\ref{alg:flitsr} is
indeed minimal, i.e., a basis. The proof is given in the replication package.

%% file: motiv.tex
\subsection{Running Example}
%
\label{sec:motv}
We use the code in Listing~\ref{lst:motv1} to
illustrate our approach.  The \texttt{count} method returns a
three-element array \texttt{cs} containing the number of characters in the input string
\texttt{s} that fall within three different character classes (i.e., upper-
and lower-case letters, and digits), or \texttt{null} if it encounters any
other character. The \texttt{type} method checks whether the string
corresponding to the counts \texttt{cs} is a number or a word, or neither. This
code contains multiple faults; for now we focus on those
on lines 6, 9, and 23.
%

\begin{lstlisting}[float=t,mathescape, language=java, label={lst:motv1},
caption={Code example with multiple faults; locations and fixes indicated by comments.}, float]
public int[] count(String s) {
  int len = s.length();
  int[] cs = new int[3]; // [upper,lower,digits]
  for (int i = 0; i < len; i++) {
    if (48 <= s.charAt(i) && s.charAt(i) <= 57) {
      cs[len-1]++; // fix: cs[cs.length-1]++
    } else if (65 <= s.charAt(i) && s.charAt(i) <= 90) {
      cs[0]++;
    } else if (96 <= s.charAt(i) && s.charAt(i) <= 120) {         // fix:$\:$ (97 <= s.charAt(i) && s.charAt(i) <= 122)
      cs[1]++;
    } else {
      return null;
    }
  }
  return cs;
}

public String type(int[] cs) {
  if (cs == null) {
    return "invalid input";
  }
  int len = cs[0] + cs[1] + cs[2];
  if (cs[len-1] == len) { // fix: cs[cs.length-1]
    return "number";
  } else if (cs[0] + cs[1] == len) {
    return "word";
  }
  return "unknown";
}
\end{lstlisting}

\Cref{tab:motv1} shows
the test suite, the spectrum, and
the suspiciousness scores computed for the example
metrics.
We use a test suite of
15~unit tests for
\texttt{count} ($c_{1}$--$c_{15}$), 8~unit tests for \texttt{type} ($t_{16}$--$t_{23}$), and
3~integration tests ($i_{24}$--$i_{26}$). The integration tests call \texttt{type} on the
result returned by \texttt{count} for the given input strings.
Note that SBFL can only localize faults that are
executed in at least one failing test case, and that the test suite ensures this for these three faults.
We use
Tarantula~\cite{tarantula}, Ochiai~\cite{ochiai},
and DStar~\cite{dstar} as example metrics.

The metrics all
agree on $l_{12}$ as the most suspicious statement; unfortunately, this is not
faulty.  The reasons for this confident but wrong verdict are that
there are only three tests (i.e., $c_{3}$, $c_{4}$, and $c_{5}$) that execute
$l_{12}$, which all fail, giving it a high score under any metric,
but the control flow fault at $l_{9}$ forces the execution of $l_{12}$ in
these failing tests, so that $l_{12}$ masks the actual fault location $l_{9}$.
Even so, Tarantula performs relatively well, ranking the faults in second, third,
and tied fourth position, respectively, which gives an
AWE$_L$
of 1.5 statements while
Ochiai and DStar perform worse, with larger AWEs of 6.5 and 8, respectively.
We show
below how FLITSR improves on these results.

%

\begin{figure}
\setlength{\tabcolsep}{1pt}
\resizebox{\columnwidth}{!}{%
  \begin{tabular}{|c|c|c|c|c|c|c|c|c|c|c|c|c|c|c|c|c|c|c|c|c|}\hline
    \# & input & $l_2$ & $l_3$ & $l_4$ & $l_5$ & \textbf{\emph{l}}$_\mathbf{6}$ & $l_7$
               & $l_8$ & \textbf{\emph{l}}$_\mathbf{9}$ & $l_{10}$ & $l_{12}$
	       & $l_{15}$ & $l_{19}$ & $l_{20}$ & $l_{22}$
	       & \textbf{\emph{l}}$_\mathbf{23}$ & $l_{24}$ & $l_{25}$ & $l_{26}$ & $l_{28}$\\
    \hline
    \rowcolor{red!20}
    $c_{1}$ & \texttt{"0"}      &\f&\f&\f&\f&\f&&&&&&\f&&&&&&&&\\
    \rowcolor{red!20}
    $c_{2}$ & \texttt{"\textasciigrave{}bc"} &\f&\f&\f&\f&&\f&&\f&\f&&\f&&&&&&&&\\
    \rowcolor{red!20}
    $c_{3}$ & \texttt{"az"}     &\f&\f&\f&\f&&\f&&\f&&\f&&&&&&&&&\\
    \rowcolor{red!20}
    $c_{4}$ & \texttt{"1z"}     &\f&\f&\f&\f&\f&\f&&\f&&\f&&&&&&&&&\\
    \rowcolor{red!20}
    $c_{5}$ & \texttt{"Ay"}     &\f&\f&\f&\f&&\f&\f&\f&&\f&&&&&&&&&\\
    $c_{6}$ & \texttt{"AZ"}     &\p&\p&\p&\p&&\p&\p&&&&\p&&&&&&&&\\
    $c_{7}$ & \texttt{"bg"}     &\p&\p&\p&\p&&\p&&\p&\p&&\p&&&&&&&&\\
    $c_{8}$ & \texttt{"Bg"}     &\p&\p&\p&\p&&\p&\p&\p&\p&&\p&&&&&&&&\\
    $c_{9}$ & \texttt{"gx"}     &\p&\p&\p&\p&&\p&&\p&\p&&\p&&&&&&&&\\
    $c_{10}$ & \texttt{"x"}     &\p&\p&\p&\p&&\p&&\p&\p&&\p&&&&&&&&\\
    $c_{11}$ & \texttt{"AGMZ"}  &\p&\p&\p&\p&&\p&\p&&&&\p&&&&&&&&\\
    $c_{12}$ & \texttt{"A"}     &\p&\p&\p&\p&&\p&\p&&&&\p&&&&&&&&\\
    $c_{13}$ & \texttt{"a"}     &\p&\p&\p&\p&&\p&&\p&\p&&\p&&&&&&&&\\
    $c_{14}$ & \texttt{"ZZ"}    &\p&\p&\p&\p&&\p&\p&&&&\p&&&&&&&&\\
    $c_{15}$ & \texttt{"abcde"} &\p&\p&\p&\p&&\p&&\p&\p&&\p&&&&&&&&\\\hline
    \rowcolor{red!20}
    $t_{16}$ & \texttt{[0,2,0]} &&&&&&&&&&&&\f&&\f&\f&\f&&&\\
    \rowcolor{red!20}
    $t_{17}$ & \texttt{[0,0,1]} &&&&&&&&&&&&\f&&\f&\f&&\f&&\f\\
    \rowcolor{red!20}
    $t_{18}$ & \texttt{[0,0,6]} &&&&&&&&&&&&\f&&\f&\f&&&&\\
    $t_{19}$ & \texttt{[0,1,1]} &&&&&&&&&&&&\p&&\p&\p&&\p&&\p\\
    $t_{20}$ & \texttt{[0,0,3]} &&&&&&&&&&&&\p&&\p&\p&\p&&&\\
    $t_{21}$ & \texttt{[1,0,1]} &&&&&&&&&&&&\p&&\p&\p&&\p&&\p\\
    $t_{22}$ & \texttt{[1,1,1]} &&&&&&&&&&&&\p&&\p&\p&&\p&&\p\\
    $t_{23}$ & \texttt{null}    &&&&&&&&&&&&\p&\p&&&&&&\\\hline
    $i_{24}$ & \texttt{"aZ"}    &\p&\p&\p&\p&&\p&\p&\p&\p&&\p&\p&&\p&\p&&\p&\p&\\
    $i_{25}$ & \texttt{"91"}    &\p&\p&\p&\p&\p&&&&&&\p&\p&&\p&\p&\p&&&\\
    $i_{26}$ & \texttt{"000"}   &\p&\p&\p&\p&\p&&&&&&\p&\p&&\p&\p&\p&&&\\
    \hline
    \hline
    \multicolumn{2}{|l|}{~Tarantula} &0.46&0.46&0.46&0.46&0.69&0.45&0.27&0.56&0.24&\tp 1.00&0.26&0.46&0.00&0.49&0.49&0.43&0.36&0.00&0.43\\
    \multicolumn{2}{|l|}{~Ochiai} &0.42&0.42&0.42&0.42&0.35&0.37&0.13&0.43&0.13&\tp 0.61&0.18&0.32&0.00&0.34&0.34&0.18&0.16&0.00&0.18\\
    \multicolumn{2}{|l|}{~DStar} &1.56&1.56&1.56&1.56&0.50&1.07&0.08&1.45&0.07&\tp 1.80&0.21&0.69&0.00&0.75&0.75&0.10&0.09&0.00&0.10\\\hline
  \end{tabular}
}
\caption{Test suite, spectrum, and suspiciousness scores for running example.
\p\ and \f\ denote execution in passing and failing (also highlighted in red) tests.
Faulty statements are boldfaced in the top row, scores of the most
suspicious statements are highlighted in yellow in the bottom rows.
}
\label{tab:motv1}
%
\end{figure}

%% file: mult.tex
\section{Multi-Round Localization}%
%
\label{sec:multi}

The FLITSR core algorithm computes a single basis comprising only
a few, highly suspicious elements. However, most program spectra admit multiple bases, with each
giving an alternate explanation for the test failures.
In this section, we describe a multi-round extension called FLITSR* which
iteratively calculates additional bases, and ranks more program elements explicitly.
We illustrate this extension with an extension of our running example, as shown in
Figure~\ref{tab:motv2} and described below.

FLITSR* is particularly
useful when the program contains dominators and 
more than one failure-explaining basis may be required.
More specifically, since a dominator, by definition
(see~\Cref{sec:background}), masks \emph{all} other
exposed faults that it dominates, FLITSR may fail to localize
them; this occurs when the dominator is ranked by the base metric as the most suspicious
element in any iteration (but the last) of the first phase, since FLITSR either removes
all (remaining) failing tests executing the dominator and thus does not consider the masked faults
in the first phase, or removes them from the basis in the second phase if they
had already been selected.

\mysubsubsection{Running example}
Consider again our running example in \Cref{lst:motv1} and~\Cref{tab:motv1}.
$l_2$ is actually faulty as well and fails
with an uncaught exception if \texttt{count} is called with a \texttt{null}
value. However, the original test suite
does not expose this fault, and the base
metrics score $l_2$ fairly low.
%
If we now add to $T$
the new integration test $i_{27}$
that calls \texttt{count} with
\texttt{null} (cf.\ \Cref{tab:motv2}),
the base metrics score $l_2$ higher.

\begin{figure}
\setlength{\tabcolsep}{1pt}
\resizebox{\columnwidth}{!}{%
  \begin{tabular}{|l|c|c|c|c|c|c|c|c|c|c|c|c|c|c|c|c|c|c|c|c|}\hline
	  \# & input & \textbf{\emph{l}}$_\mathbf{2}$ & $l_3$ & $l_4$ & $l_5$ & \textbf{\emph{l}}$_\mathbf{6}$ & $l_7$
               & $l_8$ & \textbf{\emph{l}}$_\mathbf{9}$ & $l_{10}$ & $l_{12}$
	       & $l_{15}$ & $l_{19}$ & $l_{20}$ & $l_{22}$
	       & \textbf{\emph{l}}$_\mathbf{23}$ & $l_{24}$ & $l_{25}$ & $l_{26}$ & $l_{28}$\\
    \hline
    \rowcolor{red!20}
    $c_{1}$ & \texttt{"0"}      &\f&\f&\f&\f&\f&&&&&&\f&&&&&&&&\\
    \multicolumn{1}{|c|}{\vdots} &
    \multicolumn{1}{c|}{\vdots} &
    \multicolumn{1}{c|}{\vdots} &
    \multicolumn{1}{c|}{\vdots} &
    \multicolumn{1}{c|}{\vdots} &
    \multicolumn{1}{c|}{\vdots} &
    \multicolumn{1}{c|}{\vdots} &
    \multicolumn{1}{c|}{\vdots} &
    \multicolumn{1}{c|}{\vdots} &
    \multicolumn{1}{c|}{\vdots} &
    \multicolumn{1}{c|}{\vdots} &
    \multicolumn{1}{c|}{\vdots} &
    \multicolumn{1}{c|}{\vdots} &
    \multicolumn{1}{c|}{\vdots} &
    \multicolumn{1}{c|}{\vdots} &
    \multicolumn{1}{c|}{\vdots} &
    \multicolumn{1}{c|}{\vdots} &
    \multicolumn{1}{c|}{\vdots} &
    \multicolumn{1}{c|}{\vdots} &
    \multicolumn{1}{c|}{\vdots} &
    \multicolumn{1}{c|}{\vdots}
    \\
    $i_{26}$ & \texttt{"000"}   &\p&\p&\p&\p&\p&&&&&&\p&\p&&\p&\p&\p&&&\\
    \rowcolor{red!20}
    t27 & \texttt{null} &\f&&&&&&&&&&&&&&&&&&\\
    \hline
    \hline
    \multicolumn{2}{|l|}{Tarantula} &0.48&0.44&0.44&0.44&0.67&0.42&0.25&0.53&0.22&\tp 1.00&0.24&0.43&0.00&0.46&0.46&0.40&0.33&0.00&0.40\\
    \multicolumn{2}{|l|}{Ochiai}&0.46&0.39&0.39&0.39&0.33&0.34&0.13&0.40&0.12&\tp 0.58&0.17&0.30&0.00&0.32&0.32&0.17&0.15&0.00&0.17\\
    \multicolumn{2}{|l|}{DStar}&\tp 2.25&1.47&1.47&1.47&0.44&1.00&0.07&1.33&0.07&1.50&0.20&0.64&0.00&0.69&0.69&0.09&0.08&0.00&0.09\\
    \hline \hline
    \multicolumn{2}{|l|}{1.1}&0.46&0.39&0.39&0.39&0.33&0.34&0.13&0.40&0.12&\tp0.58&0.17&0.30&0.00&0.32&0.32&0.17&0.15&0.00&0.17\\
    \multicolumn{2}{|l|}{1.2}&0.31&0.21&0.21&0.21&0.24&0.12&0.00&0.14&0.14&-&0.21&0.37&0.00&\tp\textbf{0.39}&\tp\textbf{0.39}&0.20&0.18&0.00&0.20\\
    \multicolumn{2}{|l|}{1.3}&\tp\textbf{0.43}&0.30&0.30&0.30&0.33&0.17&0.00&0.20&0.20&-&0.30&0.00&0.00&-&-&0.00&0.00&0.00&0.00\\
    \hline
    \multicolumn{2}{|l|}{FLITSR} &\tp \#2&-&-&-&-&-&-&-&-&-&-&-&-&\tp \#1&\tp \#1&-&-&-&-\\
    \hline \hline
    \multicolumn{2}{|l|}{2.1}&-&0.42&0.42&0.42&0.35&0.37&0.13&0.43&0.13&\tp0.61&0.18&0.32&0.00&-&-&0.18&0.16&0.00&0.18\\
    \multicolumn{2}{|l|}{2.2}&-&0.23&0.23&0.23&0.26&0.13&0.00&0.16&0.16&-&0.23&\tp\textbf{0.40}&0.00&-&-&0.22&0.20&0.00&0.22\\
    \multicolumn{2}{|l|}{2.3}&-&0.37&0.37&0.37&\tp\textbf{0.41}&0.20&0.00&0.25&0.25&-&0.37&-&0.00&-&-&0.00&0.00&0.00&0.00\\
    \multicolumn{2}{|l|}{2.4}&-&0.27&0.27&0.27&-&0.29&0.00&\tp\textbf{0.35}&\tp0.35&-&0.27&-&0.00&-&-&0.00&0.00&0.00&0.00\\
    \hline
    \multicolumn{2}{|l|}{3.1}&-&0.45&0.45&0.45&-&0.39&0.14&-&0.13&0.65&0.20&-&0.00&-&-&0.19&0.17&0.00&0.19\\
    \multicolumn{2}{|l|}{3.2}&-&\textbf{0.26}&\textbf{0.26}&\textbf{0.26}&-&0.14&0.00&-&0.18&-&0.26&-&0.00&-&-&0.25&0.22&0.00&0.25\\
    \multicolumn{2}{|l|}{3.3}&-&-&-&-&-&0.00&0.00&-&0.00&-&0.00&-&0.00&-&-&\textbf{0.35}&0.32&0.00&0.35\\
    \multicolumn{2}{|l|}{3.4}&-&-&-&-&-&0.00&0.00&-&0.00&-&0.00&-&0.00&-&-&-&0.45&0.00&\textbf{0.50}\\
    \hline
    \multicolumn{2}{|l|}{4.1}&-&-&-&-&-&0.42&0.15&-&0.14&0.71&0.21&-&0.00&-&-&-&0.18&0.00&-\\
    \multicolumn{2}{|l|}{4.2}&-&-&-&-&-&0.17&0.00&-&0.20&-&\textbf{0.30}&-&0.00&-&-&-&0.26&0.00&-\\
    \multicolumn{2}{|l|}{4.3}&-&-&-&-&-&0.00&0.00&-&0.00&-&-&-&0.00&-&-&-&\textbf{0.45}&0.00&-\\
    \hline
    \multicolumn{2}{|l|}{5.1}&-&-&-&-&-&\textbf{0.52}&0.19&-&0.18&-&-&-&0.00&-&-&-&-&0.00&-\\
    \hline
    \multicolumn{2}{|l|}{6.1}&-&-&-&-&-&-&\textbf{0.27}&-&0.25&-&-&-&0.00&-&-&-&-&0.00&-\\
    \multicolumn{2}{|l|}{6.2}&-&-&-&-&-&-&-&-&\textbf{0.35}&-&-&-&0.00&-&-&-&-&0.00&-\\
    \hline
    \multicolumn{2}{|l|}{FLITSR$^*$}
    &\#2&\#6&\#6&\#6&\#4&\#12&\#13&\#5&\#14&\#9&\#10&\#3&-&\#1&\#1&\#7&\#11&-&\#8\\
    \hline
  \end{tabular}
}
\caption{Spectrum (top block) and suspiciousness scores for running
	example with extended test suite. 
	Scores for base metrics shown in second block, Ochiai scores for FLITSR and FLITSR* rounds 
	in third and fourth blocks. 
	Highest scores in each iteration are highlighted, and elements in any computed basis are boldfaced. Final ranking shown at bottom.
}
\label{tab:motv2}
\end{figure}

However, if we run FLITSR using this extended test suite $T'$, we encounter $l_2$ early on, and since it is a
dominator for \texttt{count}, FLITSR constructs an ``impoverished'' basis, which
leaves too many (faulty) elements unranked. Consider for example FLITSR's run
using Ochiai, as detailed in the lower part of \Cref{tab:motv2}.
As before (i.e., without $i_{27}$), FLITSR first selects $l_{12}$, then
$l_{22}$ and $l_{23}$ in the first two iterations of its first phase, removing
the corresponding test failures $\{c_3, c_4, c_5, t_{16},t_{17}, t_{18}\}$
from
$T'$.
In the third iteration, it identifies $l_2$ as most suspicious, due to the
failing test case $i_{27}$; however, since this is a dominator for
\texttt{count}, FLITSR removes all remaining test failures $c_1$, $c_2$, and $i_{27}$ and
terminates its first phase. In the second phase, it identifies $l_{12}$ as
redundant.  The basis returned by FLITSR is thus just
$\{l_2, l_{22}, l_{23}\}$, with $l_2$ at rank \#2, as given in~\Cref{tab:motv2}.
%
Hence, FLITSR fails to localize the faults $l_6$ and $l_9$
due to the presence of the exposed dominator and its performance degrades.

\mysubsubsection{FLITSR*}
We can prevent this performance degradation by running multiple FLITSR
rounds over increasingly smaller subsets of the SUT's elements and test suite.
Each round then returns a less suspicious basis for the remaining test
failures. The elements in each round's basis $C_i$ are ranked just below those
of the previous round.

We remove the basis $C_i$ before the next round, i.e., do not compute any
scores for these elements. We also remove the failing tests that are executed
\emph{only} by elements in $C_i$ to prevent these now ``unexplainable''
test failures from influencing the scores computed in the next round.
We repeat this process until a round can no longer compute a basis,
i.e., until all elements executing a failing test have been added to a basis
and thus been ranked explicitly.

\begin{algorithm}[t]
  \caption{FLITSR$^*$ multi-round algorithm}\label{alg:multi}
{\small
  \begin{algorithmic}[1]
    \Require{$M$}                     \Comment{base scoring method}
    \Require{$S_0$}                   \Comment{system under test}
    \Require{$T_0$}                   \Comment{initial test suite}
    \Ensure{$C^*$}                    \Comment{basis list}
    \Let{$C^*$}{$\langle\rangle$}
    \State{{\sc flitsr$^*$}($S_0, T_0$)}
  \vspace*{0.25ex}
  \Function{\sc flitsr*}{$S_i$, $T_i$}
    \Let{$C$, \_}{{\sc flitsr}($M, S_i, T_i$)}
      \Comment{Algorithm~\ref{alg:flitsr}}
    \If{$C\neq\emptyset$}
      \Let{\parbox{1em}{$C^*$}}{$C^*+\!\!\!+\,\langle C\rangle$} \Comment{append basis with lowest rank}
      \Let{\parbox{1em}{$S'_i$}}{$S_i \setminus C$}              \Comment{remove from coverage matrix}
      \Let{\parbox{1em}{$R$}}{$F_{S_i,T_i}(C) \setminus F_{S_i,T_i}(S'_i)$}
        \Comment{tests failing in basis only}
      \Let{\parbox{1em}{$T'_i$}}{$T_i \setminus R$}              \Comment{remove from coverage matrix}
      \State{{\sc flitsr$^*$}($S'_i, T'_i$)}
    \EndIf
    \EndFunction
  \end{algorithmic}
}
\end{algorithm}

Algorithm~\ref{alg:multi} shows the algorithm's details. It reuses
Algorithm~\ref{alg:flitsr} with minor changes as basic building block.  In
particular, we now require the algorithm to maintain and use an offset in the
rank assignments, to ensure that later bases are ranked lower.
%
%
%
%

\mysubsubsection{Running example}
The lower part of \Cref{tab:motv2} shows the detailed Ochiai scores in the
subsequent rounds of FLITSR$^*$ for our running example.  Entries dashed across
the entire block denote the elements removed in a previous round.

In the first round, FLITSR identifies the basis $C_1=\{l_2, l_{22}, l_{23}\}$.
These elements (in fact, $l_2$) exclusively execute the failing test $t_{27}$,
so for the second round we remove $C_1$ from $S$ and $F_{S,T}(C_1)\setminus
F_{S,T}(S\backslash C_1) = \{t_{27}\}$ from $T$, for the new SUT $S_1$ and test
suite $T_1$, respectively.
This second round identifies the basis $C_2=\{l_6, l_9, l_{19}\}$. All failing
tests remaining in $T_1$ that are executed by elements in $C_2$ are also
executed by elements remaining in $S_1$ that are not in $C_2$, so we start the
third round with $S_1\setminus C_2$ and $T_1$. The following rounds identify the
elements shown in bold in the table as basis and remove them and the associated
failing test.
After the sixth round no failing tests remain, and FLITSR returns an empty basis,
so we terminate the (outer) FLITSR$^*$ loop.

Using the ranking in the last line of \Cref{tab:motv2} gives us an AWE$_L$
of 2, which is a substantial improvement over both FLITSR's
single-round result of 12.33, and the Ochiai result of 5.

%% file: eval.tex
\section{Experimental Evaluation}
%
\label{sec:eval}

\subsection{Research Questions}
%
\label{sec:eval:rqs}
With our experimental evaluation, we aim to answer the following
research questions:
\begin{description}
  \item[RQ1] How much does FLITSR improve on existing SBFL methods?
  \item[RQ2] How much are the results dependent on the applied base metric?
    Which base metric, if any, induces the best results?
  \item[RQ3] How much does FLITSR improve on learning-based methods?
  \item[RQ4] What effect does the multi-round variant FLITSR$^*$ have
    on the improvements?
  \item[RQ5] What effect do the fault type (real or
    injected) and coverage granularity have on the results of FLITSR and FLITSR*?
\end{description}



\subsection{Experimental Setup}
%
\label{sec:eval:setup}
%

\subsubsection{Synthetic Fault Data Set}
\label{sec:eval:steimann:setup}
%
Our first experimental dataset is based on variants of 15 Java-based
open-source systems%
\footnote{Eclipse Draw2d, Eventbus, Jester, JExel, JParsec, Jaxen, HTML Parser,
Apache Commons Codec, Apache Commons Lang, Daikon, Barbecue, JDepend, Mime4J,
Time \& Money, XMLSec.}
analyzed by
Steimann and Frenkel \cite{parallel:steimann,tcm}.
Each variant differs from its respective baseline system
by a number of \emph{synthetic faults injected} into its methods by repeatedly
applying simple mutation operators.  The number $N_f$ of injected faults follows
a geometric progression; the number of resulting variants is $m_1\leq 600$
resp.\ $m_n=5000, n=2,4,8,16,32$, for a total of 75000 $n$-fault
variants.

Each variant was executed over its respective baseline test suite, which
range in size from 55 (JDepend) to 1666 (Daikon) tests ($\mu\!\:=\!\:345,
\sigma^2\!\:=\!\:403$). However, not all methods are exercised by the test suites,
and the spectra contain between 152 (Jester) and 2075 (Lang) methods
($\mu\!\:=\!\:698, \sigma^2\!\:=\!\:589$).

\subsubsection{Real Fault Data Set}
\label{sec:eval:d4j:setup}
%

\begin{table}
  \caption{Defects4J (1.5.0) multi-fault benchmark characteristics;
	   cf.\ \cite{multid4j} for included bug ids. $N$ is the number of versions in each project.}
  \centering
  {\small
   \setlength{\tabcolsep}{1.5pt}
   \setlength{\doublerulesep}{0.75pt}
  \begin{tabular}{|l|r||rr|rr|rr|rr|lcr|}\hline
	  &     & \multicolumn{4}{c|}{Size (methods / stmts.)} & \multicolumn{4}{c|}{Tests (pass / fail)} & \multicolumn{3}{c|}{Faults} \\
  Project & $N$ &  min &  max &   min &   max &  min &  max & min & max & min &  avg & max             \\
  \hline
  Chart   &  26 & 3466 & 4135 & 62113 & 79788 & 1578 & 6486 &   1 &  24 &   1 & 3.39 &  5
  \\
  Closure & 104 & 3294 & 5520 & 27412 & 45109 & 2591 & 7821 &   7 &  81 &   2 & 9.10 & 14
  \\
  Lang    &  64 & 1538 & 2022 &  8909 & 11190 & 1593 & 2658 &   2 &  35 &   2 & 6.10 & 12
  \\
  Math    & 106 & 745  & 4029 &  4809 & 38688 &  876 & 5184 &   2 &  34 &   1 & 4.90 & 10
  \\
  Time    &  26 & 1732 & 1797 &  9898 & 10917 & 3745 & 3996 &   6 &  54 &   2 & 6.94 & 10
  \\\hline
  \end{tabular}
  }
  \label{tab:d4j-benchmarks}
\end{table}

Our second experimental dataset is based on Defects4J (v1.5.0)
\cite{defects4j}, a collection of real-world faults extracted from five medium-sized
(20--100 kLoC) Java-based open-source systems.  Each fault is given in the form
of a diff file that repairs the faulty base version, and is accompanied by a test
suite that exposes this fault only.
The variants in Defects4J are not necessarily single-fault systems,
since faults from later versions can already be present in earlier ones. However these
multiple faults are not identified in each version since the test cases that
expose the faults do not initially exist in that version.
An et al.~\cite{multid4j} use an iterative search approach to ``transplant''
the fault-exposing tests across versions in order to \emph{identify} versions
within the Defects4J dataset that \emph{already} contain multiple faults. From
this, they detect 311 existing multi-fault versions within the Defects4J dataset.

However, the test transplantation by An et al.\cite{multid4j} does not
provide the localization oracle that is required to evaluate the performance of
a fault localizer:
it only guarantees that tests that expose a
fault $f_i$ in a more recent system variant $v_i$ also fail in the older
variant $v_j$, but does not specify where in $v_j$ the fault $f_i$ is located.
%

We therefore apply a \emph{fault location translation}
process that augments the test transplantation. For this process, we identify
all lines in the diff that repairs $f_i$ in $v_i$ as faulty lines pertaining to
that bug. We then locate these lines in the earlier version $v_j$ (where the
lines may have been moved), by considering the diffs in the full project history
in order, updating the line numbers if their position has changed.
If a line is either not executable, or modified during the history of the project
between the versions, we consider the line no longer necessarily identifying for
the fault, and therefore remove it from consideration. If all the lines in a
fault-fixing diff have been dropped in this way, a characteristic location for the
fault cannot be given, and we therefore remove this fault from the experimental
evaluation. In this way, we get a subset of \emph{all} faults identified by
An et al.~\cite{multid4j} in \emph{each} version.
Note that we do not alter any Defects4J version, but merely identify the
statements in a version that pertain to a bug that has been described in a later
version.
This is in contrast to Zheng et al.~\cite{DBLP:journals/jss/ZhengWFCY18}, who
\emph{created} synthetic multi-fault versions by transplanting \emph{faults}.
\Cref{tab:d4j-benchmarks} summarizes our dataset; it is available at 
\url{https://github.com/DCallaz/defects4j_multifault}.



\subsubsection{Data Collection}
\label{sec:eval:steimann:setup}

For the first dataset, we used the method-level execution spectra provided by
Steimann and Frenkel \cite{parallel:steimann,tcm},
%
%
in the form of TCMs.
For the second dataset, we compiled the (faulty) sources using the Defect4J
compile script and relied on GZoltar (v1.7.3)~\cite{gzoltar} to collect
coverage information for each of the variants. We condensed this into method-
and statement-level spectra.

For each spectrum we computed eight single fault metrics
(Tarantula, Ochiai, DStar, Jaccard, GP13, naish2, Overlap, and Harmonic) and
three multi-fault metrics (Zoltar, Hyperbolic and Barinel), and ran both FLITSR
and FLITSR* using each of these as base metric. For each metric (including
all FLITSR variants) we produced a single suspiciousness ranking per spectrum. We
evaluated the metrics based on the rankings without attempting to fix any
faults.

We also compare to the
state-of-the-art learning-based tool GRACE, which has been shown to outperform other
learning-based techniques~\cite{grace}. For this comparison, we ran the existing
implementation of GRACE (which identifies method-level faults only) as a black-box over
the five Defects4J projects we use in our evaluation. Note that the actual
Defects4J code bases are \emph{identical} to those used for the original
evaluation of GRACE~\cite{grace}, with the only difference being the test suite
supplied for each version.
%
%
Note also that FLITSR runs almost instantaneously, once the coverage data is
collected, while GRACE spent several hours on optimizing its predictions, and
failed to terminate within 24~hours for Closure.

\subsection{Method-Level Localization Results}
\label{sec:eval:steimann}

\label{sec:eval:steimann:results}
%

\begin{table*}
  \caption{
    Method-level results for 15 Java projects \cite{parallel:steimann}
    averaged over all variants with given number $N_f$ of mutations.
    Metrics designed for multi-fault localization are shown in the last three rows.
    Improvements of FLITSR and FLITSR* variants over corresponding base metrics are
    highlighted in yellow.  Best results in each column are shown in bold.
  }
  \vspace*{-0.5ex}
  \centering
  {\small
   \setlength{\tabcolsep}{1.2pt}
   \setlength{\doublerulesep}{0.5pt}
   \begin{tabular}{|l||r|r|r||r|r|r|r||r|r|r|r||r|r|r|r||r|r|r|r||r|r|r|r|}\hline
    & \multicolumn{3}{c||}{$N_f=1$}
    & \multicolumn{4}{c||}{$N_f=2$}
    & \multicolumn{4}{c||}{$N_f=4$}
    & \multicolumn{4}{c||}{$N_f=8$}
    & \multicolumn{4}{c||}{$N_f=16$}
    & \multicolumn{4}{c|}{$N_f=32$}\\
    \hline
    Metric
    &{\footnotesize AWE$_1$}                        &{\footnotesize P{\scriptsize @}5}&{\footnotesize$\:\!\!$R{\scriptsize$\:\!\!$@$\:\!\!$}$N_{\!f\!}$}
    &{\footnotesize AWE$_1$}&{\footnotesize AWE$_2$}&{\footnotesize P{\scriptsize @}5}&{\footnotesize$\:\!\!$R{\scriptsize$\:\!\!$@$\:\!\!$}$N_{\!f\!}$}
    &{\footnotesize AWE$_1$}&{\footnotesize AWE$_M$}&{\footnotesize P{\scriptsize @}5}&{\footnotesize$\:\!\!$R{\scriptsize$\:\!\!$@$\:\!\!$}$N_{\!f\!}$}
    &{\footnotesize AWE$_1$}&{\footnotesize AWE$_M$}&{\footnotesize P{\scriptsize @}5}&{\footnotesize$\:\!\!$R{\scriptsize$\:\!\!$@$\:\!\!$}$N_{\!f\!}$}
    &{\footnotesize AWE$_1$}&{\footnotesize AWE$_M$}&{\footnotesize P{\scriptsize @}5}&{\footnotesize$\:\!\!$R{\scriptsize$\:\!\!$@$\:\!\!$}$N_{\!f\!}$}
    &{\footnotesize AWE$_1$}&{\footnotesize AWE$_M$}&{\footnotesize P{\scriptsize @}5}&{\footnotesize$\:\!\!$R{\scriptsize$\:\!\!$@$\:\!\!$}$N_{\!f\!}$}
    \\
    \hline
    FLITSR (T)
    &\tp    25.4 &\tp                8.3 &\tp     39.9
    &\tp    16.3 &\tp    65.5 &\tp     14.1 &\tp     33.1
    &\tp    10.4 &\tp    42.6 &\tp     22.6 &\tp     29.7
    &\tp     7.0 &\tp    48.3 &\tp     28.6 &\tp     28.0
    &\tp     4.9 &\tp    52.6 &\tp     31.0 &\tp     28.4
    &\tp     3.1 &\tp    52.6 &\tp     33.0 &\tp     31.2
    \\
    FLITSR (O)
    &       19.7 &                  10.3 &\tp     41.9
    &\tp\bf 12.0 &\tp    60.3 &\tp     18.1 &\tp \bf 35.6
    &\tp     8.5 &\tp    40.1 &\tp     26.4 &\tp     31.7
    &\tp     6.3 &\tp    49.1 &\tp     30.5 &\tp     29.5
    &\tp     5.0 &\tp    61.2 &\tp     31.6 &\tp     28.5
    &\tp     3.4 &\tp    65.1 &\tp     32.6 &\tp     29.7
    \\
    FLITSR (H)
    &       17.2 &                  10.4 &\tp     42.1
    &\tp    12.2 &\tp    61.9 &\tp     17.9 &\tp \bf 35.6
    &\tp     8.6 &\tp    38.9 &\tp     26.2 &\tp \bf 31.8
    &\tp     6.1 &\tp    45.5 &\tp     30.7 &\tp     29.7
    &\tp     4.5 &\tp    52.9 &\tp     32.5 &\tp     29.2
    &\tp     3.1 &\tp    55.0 &\tp \bf 33.5 &\tp     31.1
    \\
    \hline
    FLITSR* (T)
    &\tp    21.7 &\tp               10.2 &\tp     39.9
    &\tp    13.3 &\tp    54.5 &\tp     17.8 &\tp     33.2
    &\tp     9.0 &\tp    33.6 &\tp     25.1 &\tp     29.8
    &\tp     6.3 &\tp    37.9 &\tp     28.8 &\tp     28.1
    &\tp     4.4 &\tp\bf 42.8 &\tp     31.0 &\tp     28.6
    &\tp \bf 3.0 &\tp\bf 44.2 &\tp     33.0 &\tp \bf 31.5
    \\
    FLITSR* (O)
    &       19.9 &                  10.8 &\tp     41.9
    &\tp\bf 12.0 &\tp    52.0 &\tp \bf 18.9 &\tp \bf 35.6
    &\tp \bf 8.0 &\tp    32.0 &\tp \bf 26.8 &\tp \bf 31.8
    &\tp \bf 5.8 &\tp    37.4 &\tp     30.6 &\tp     29.7
    &\tp     4.7 &\tp    45.4 &\tp     31.7 &\tp     29.1
    &\tp     3.2 &\tp    48.5 &\tp     32.6 &\tp     30.8
    \\
    FLITSR* (H)
    &       17.3 &                  10.8 &\tp     42.1
    &\tp    12.1 &\tp\bf 49.2 &\tp \bf 18.9 &\tp \bf 35.6
    &\tp     8.3 &\tp\bf 31.9 &\tp \bf 26.8 &\tp \bf 31.8
    &\tp \bf 5.8 &\tp\bf 37.1 &\tp \bf 30.8 &\tp \bf 29.8
    &\tp \bf 4.3 &\tp    43.2 &\tp \bf 32.6 &\tp \bf 29.5
    &\tp     3.1 &\tp    45.5 &\tp \bf 33.5 &\tp \bf 31.5
    \\
    \hline
    Tarantula~\cite{tarantula}
    &    29.7 &                7.3 &     22.2
    &    23.9 &    71.6 &     10.1 &     14.3
    &    20.5 &    54.2 &     11.1 &     11.5
    &    16.7 &    64.9 &     11.4 &     11.3
    &    12.2 &    72.3 &     13.2 &     13.1
    &     7.9 &    75.9 &     17.4 &     17.3
    \\
    Ochiai~\cite{ochiai}
    &    19.7 &               10.3 &     40.5
    &    12.4 &    74.0 &     14.7 &     29.9
    &    10.1 &    64.7 &     16.6 &     21.5
    &     8.8 &    89.5 &     16.5 &     15.7
    &     8.3 &   102.8 &     16.3 &     13.2
    &     6.7 &   104.1 &     17.7 &     15.1
    \\
    DStar~\cite{dstar}
    &    19.5 &               10.6 &     41.3
    &    12.5 &    83.0 &     14.6 &     29.9
    &    11.1 &    77.5 &     15.3 &     20.1
    &    10.5 &   105.9 &     14.2 &     13.5
    &    10.2 &   114.7 &     14.0 &     11.5
    &     7.8 &   112.0 &     16.0 &     14.1
    \\
    Jaccard~\cite{jaccard}
    &    20.7 &               10.2 &     40.1
    &    12.9 &    77.7 &     14.6 &     29.7
    &    10.4 &    70.5 &     16.1 &     21.0
    &     9.2 &   100.6 &     15.5 &     14.7
    &     8.9 &   113.6 &     15.2 &     12.2
    &     7.3 &   111.8 &     16.6 &     14.2
    \\
    GP13~\cite{gp13}
    &    18.5 &           \bf 11.2 &\bf  43.5
    &    22.4 &    95.3 &     11.2 &     22.2
    &    22.8 &    88.1 &     10.3 &     13.0
    &    18.2 &   108.6 &     10.2 &     10.4
    &    13.5 &   115.0 &     11.5 &     10.4
    &     8.8 &   112.0 &     14.8 &     13.9
    \\
    Naish2~\cite{naish2}
    &\bf 15.8 &           \bf 11.2 &\bf  43.5
    &    22.2 &    92.0 &     11.2 &     22.2
    &    22.8 &    87.8 &     10.3 &     13.0
    &    18.2 &   108.6 &     10.2 &     10.4
    &    13.5 &   115.0 &     11.5 &     10.4
    &     8.8 &   112.0 &     14.8 &     13.9
    \\
    Overlap~\cite{eval}
    &    39.8 &                4.5 &     11.6
    &    29.5 &    83.9 &      7.8 &     11.4
    &    22.0 &    60.6 &     10.4 &     10.8
    &    16.8 &    67.8 &     11.2 &     11.2
    &    12.2 &    72.9 &     13.2 &     13.0
    &     7.9 &    75.9 &     17.4 &     17.3
    \\
    Harmonic~\cite{DBLP:phd/basesearch/Lee11b}
    &    17.2 &               10.4 &     40.7
    &    12.5 &    70.0 &     14.5 &     29.5
    &     9.9 &    52.6 &     17.0 &     22.1
    &     7.6 &    66.3 &     18.3 &     17.5
    &     6.6 &    76.2 &     19.2 &     15.9
    &     5.4 &    81.5 &     20.9 &     17.9
    \\\hline
    Zoltar-s~\cite{zoltarM}
    &    18.8 &               10.5 &     41.5
    &    16.8 &    71.1 &     12.7 &     24.7
    &    12.7 &    52.4 &     14.9 &     19.1
    &     8.8 &    64.2 &     16.8 &     16.4
    &     6.3 &    72.4 &     19.5 &     16.4
    &     4.6 &    77.3 &     22.2 &     18.9
    \\
    Hyperbolic~\cite{hyperbolic}
    &    18.4 &               11.1 &     43.4
    &    20.1 &    82.5 &     12.0 &     24.0
    &    17.7 &    66.2 &     13.1 &     16.7
    &    13.3 &    75.0 &     14.1 &     14.0
    &    10.2 &    79.5 &     15.7 &     13.9
    &     6.8 &    82.0 &     18.8 &     17.2
    \\
    Barinel~\cite{barinel}
    &   629.1 &                0.0 &      0.0
    &   310.5 &   518.5 &      0.4 &      0.4
    &   168.9 &   363.2 &      0.9 &      0.9
    &    84.1 &   318.0 &      2.2 &      2.2
    &    40.6 &   266.9 &      4.9 &      4.9
    &    19.4 &   221.1 &      9.9 &      9.9
    \\\hline
  \end{tabular}
  }
  \label{tab:tcm-benchmarks}
\end{table*}

\begin{figure}
  \hspace*{-1.7ex}
  \includegraphics[scale=0.37]{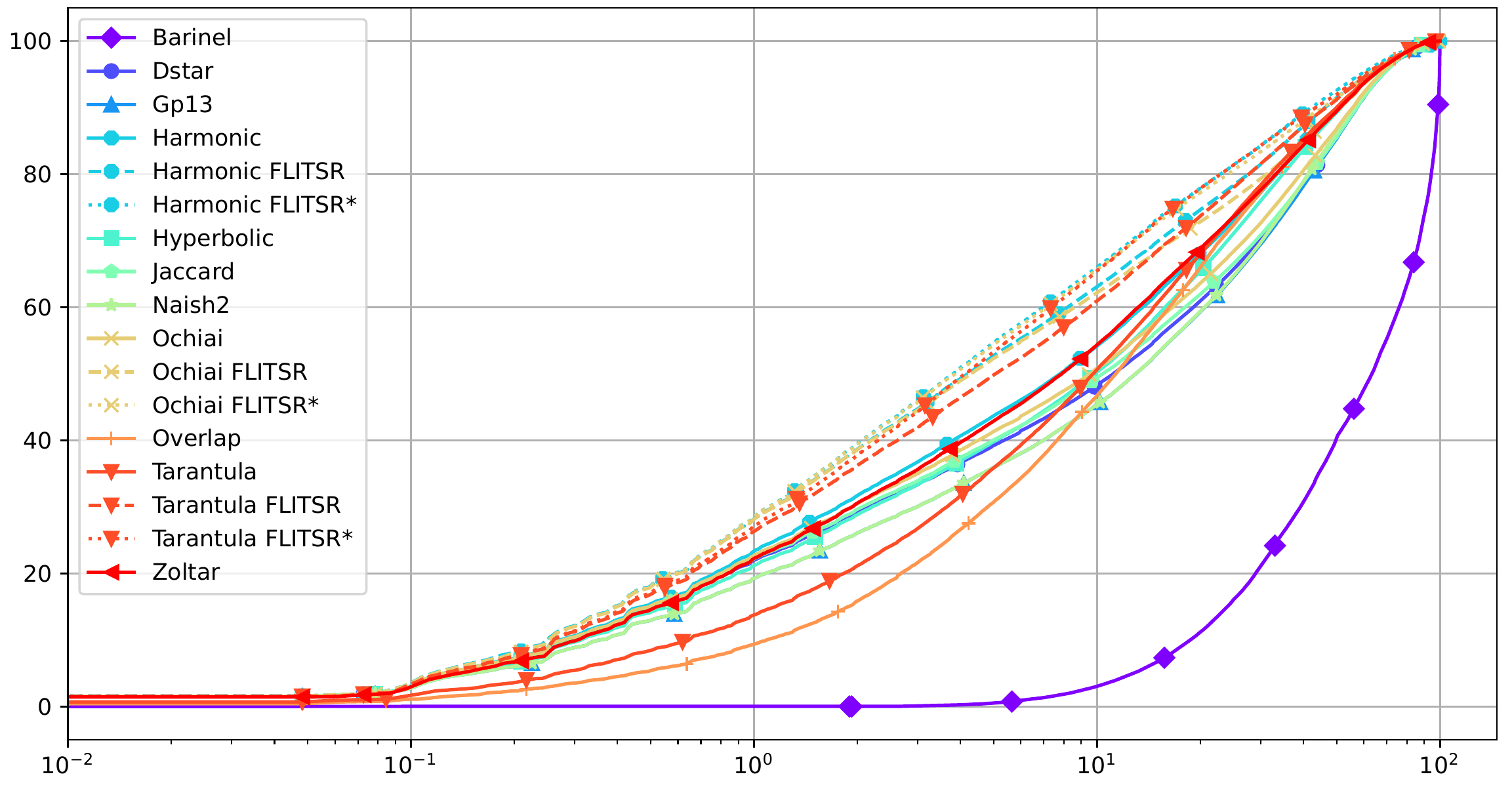}
  ~\\[-0.5ex]
  \hspace*{-1.7ex}
  \includegraphics[scale=0.37]{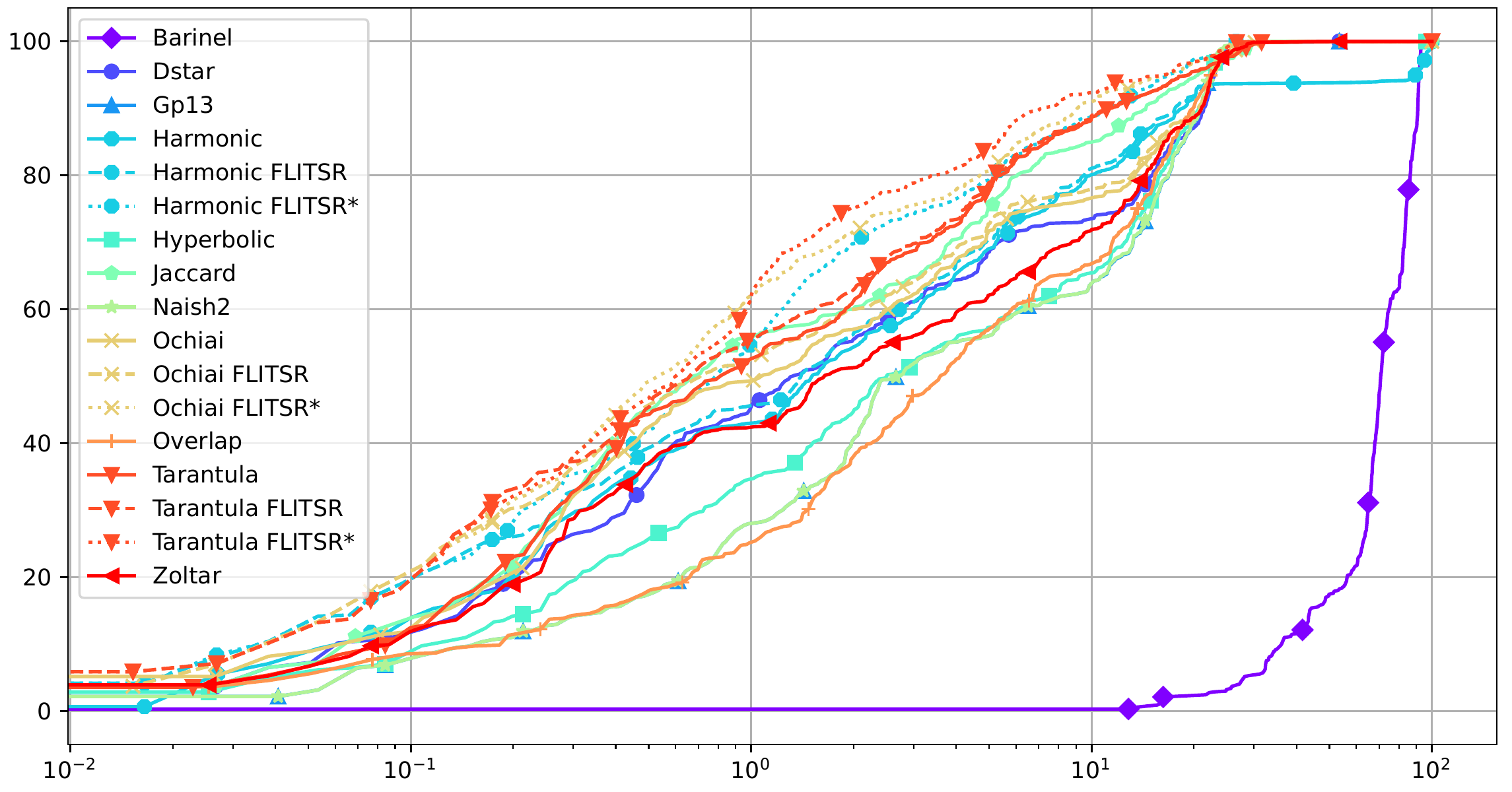}
  \caption{
    Percentage of bugs found over proportion of methods inspected in
    rank order, x-axis shown in log-scale.
    (Top) averaged over synthetic fault dataset \cite{parallel:steimann}.
    (Bot) averaged over Defects4J multi-fault dataset.
  }
  \vspace*{-2.0ex}
  \label{fig:tcm-summary}
\end{figure}

The graphs in
\Cref{fig:tcm-summary} plot the percentage of faults found with respect to the
percentage of methods inspected, over all projects in each of the datasets
and for all base metrics and a subset of FLITSR and FLITSR* variants.
We use a log-scale for the x-axis to highlight the differences at the higher
parts of the ranking since developers only inspect a small number of the
program elements in order \cite{DBLP:conf/issta/ParninO11}.
We only plot a sample of the FLITSR and FLITSR* variants because there is
little variance in the remaining results. Specifically, we show results for
Tarantula and Ochiai, due to their popularity in the literature, and
Harmonic mean, since its FLITSR and FLITSR* variants performed well.

\Cref{fig:tcm-summary} shows that the FLITSR variants find a larger
fraction of the faults at a smaller fraction of the methods inspected than
the corresponding base metrics, and that FLITSR* amplifies this trend and outperforms FLITSR
after the inspection of 0.5--1\% of the methods.
We also note that even for base metrics that perform quite differently
(cf.\ \Cref{fig:tcm-summary} for Ochiai versus Tarantula), their FLITSR and FLITSR* variants
perform surprisingly similar.
A comparison of the two graphs shows that all techniques perform substantially
better for real faults than for synthetic faults; for example, at 1\% of the
methods inspected FLITSR* already identifies about 62\% of the real faults,
compared to about 28\% of the synthetic faults.

Tables \ref{tab:tcm-benchmarks} and \ref{tab:d4j-method-results} show
more detailed results.  We include the AWEs
at different levels, as appropriate, as well as precision and
recall (cf.\ \Cref{sec:background} for definitions).
For the TCM dataset, we aggregated the data over the different fault
injection bins. Here, we use $R@N_f$, where $N_f$ is the number of injected
faults, and $P@5$.
For the Defects4J dataset, we aggregated the data over the individual
projects. Here, we always use $R@10$ (because the average number $\mean{N}{\!_f}$ of faults
per variant is less than ten), $P@5$ and $P@1$; the latter shows
how often the fault localizer pinpoints at least one fault.

\begin{table*}
  \caption{
    Method-level results for Defects4J multi-fault dataset
    averaged over all variants per project.
    See Table~\ref{tab:tcm-benchmarks} for notation.
  }
  \vspace*{-0.75ex}
  \label{tab:d4j-method-results}
  \centering
  {\footnotesize
   \setlength{\tabcolsep}{1.5pt}
   \setlength{\doublerulesep}{0.75pt}
  \begin{tabular}{|l||r|r|r|r|r||r|r|r|r|r||r|r|r|r|r||r|r|r|r|r||r|r|r|r|r|}\hline
    & \multicolumn{5}{c||}{Chart ($\mean{|S|}=3649$, $\mean{N}{\!_f}=3$)}
    & \multicolumn{5}{c||}{Closure ($\mean{|S|}=5155$, $\mean{N}{\!_f}=9$)}
    & \multicolumn{5}{c||}{Lang ($\mean{|S|}=1745$, $\mean{N}{\!_f}=6$)}
    & \multicolumn{5}{c||}{Math ($\mean{|S|}=2733$, $\mean{N}{\!_f}=5$)}
    & \multicolumn{5}{c|}{Time ($\mean{|S|}=1769$, $\mean{N}{\!_f}=7$)}\\
    \hline
    Metric
    &{\scriptsize AWE$_1$}&{\scriptsize AWE$_M$}&{\scriptsize R{\tiny @}10}&{\scriptsize P{\tiny @}1}&{\scriptsize P{\tiny @}5}
    &{\scriptsize AWE$_1$}&{\scriptsize AWE$_M$}&{\scriptsize R{\tiny @}10}&{\scriptsize P{\tiny @}1}&{\scriptsize P{\tiny @}5}
    &{\scriptsize AWE$_1$}&{\scriptsize AWE$_M$}&{\scriptsize R{\tiny @}10}&{\scriptsize P{\tiny @}1}&{\scriptsize P{\tiny @}5}
    &{\scriptsize AWE$_1$}&{\scriptsize AWE$_M$}&{\scriptsize R{\tiny @}10}&{\scriptsize P{\tiny @}1}&{\scriptsize P{\tiny @}5}
    &{\scriptsize AWE$_1$}&{\scriptsize AWE$_M$}&{\scriptsize R{\tiny @}10}&{\scriptsize P{\tiny @}1}&{\scriptsize P{\tiny @}5}  \\
    \hline
    FLITSR (T)
    &\tp \bf 0.3 &\tp \bf 4.1 &\tp \bf 78.3 &\tp \bf 67.5 &\tp     44.8
    &\tp    93.2 &      574.8 &         2.3 &\tp  \bf 3.6 &\tp      3.2
    &\tp     4.3 &       28.3 &        34.8 &\tp     44.2 &        18.5
    &\tp \bf 1.0 &\tp    12.2 &\tp     64.8 &\tp     57.1 &\tp     29.7
    &        5.2 &\tp    47.5 &\tp     16.3 &         0.0 &\tp     16.2
    \\
    FLITSR (O)
    &\tp     2.1 &\tp    22.7 &\tp     65.1 &\tp     32.5 &\tp     31.9
    &\tp    97.2 &      924.9 &\tp      2.5 &         0.0 &\tp      3.3
    &\tp \bf 3.0 &       30.5 &\tp     34.9 &\tp \bf 59.8 &\tp     21.2
    &\tp \bf 1.0 &\tp    13.2 &\tp     63.7 &\tp \bf 76.6 &\tp \bf 34.5
    &        5.1 &\tp   187.1 &        13.4 &    \bf 43.4 &        12.7
    \\
    FLITSR (H)
    &\tp     1.7 &\tp    20.8 &\tp     62.8 &\tp     41.2 &\tp     37.5
    &\tp   105.7 &      714.2 &\tp      2.4 &\tp      1.1 &\tp      3.1
    &        3.8 &       30.4 &        31.4 &\tp     45.9 &        16.7
    &\tp     1.9 &\tp    24.3 &\tp     58.9 &\tp     45.6 &\tp     26.6
    &\tp     3.0 &\tp   190.7 &        17.1 &\tp \bf 43.4 &        13.6
    \\
    \hline
    FLITSR* (T)
    &\tp \bf 0.3 &\tp     5.2 &\tp     74.7 &\tp \bf 67.5 &\tp \bf 45.3
    &\tp    50.9 &\tp\bf 520.1&\tp      2.8 &\tp  \bf 3.6 &\tp      3.3
    &\tp     3.8 &\tp\bf 21.8 &\tp \bf 36.6 &\tp     44.2 &        19.3
    &\tp \bf 1.0 &\tp\bf 12.0 &\tp     65.0 &\tp     57.1 &\tp     29.6
    &        5.2 &\tp    25.0 &\tp     16.3 &         0.0 &\tp     16.5
    \\
    FLITSR* (O)
    &\tp     2.1 &\tp     7.1 &\tp     61.5 &\tp     32.5 &\tp     32.2
    &\tp\bf 48.4 &\tp   643.3 &\tp      3.0 &         0.0 &\tp \bf  3.4
    &\tp \bf 3.0 &\tp    24.5 &\tp     36.1 &\tp \bf 59.8 &\tp \bf 22.0
    &\tp \bf 1.0 &\tp    12.5 &\tp \bf 65.2 &\tp \bf 76.6 &\tp     34.4
    &        5.1 &\tp\bf 19.3 &        16.3 &    \bf 43.4 &        12.7
    \\
    FLITSR* (H)
    &\tp     1.7 &\tp     7.0 &\tp     60.3 &\tp     41.2 &\tp     37.8
    &\tp    50.0 &\tp   652.5 &\tp  \bf 4.7 &\tp      1.1 &\tp      3.3
    &        3.8 &\tp    24.7 &\tp     32.9 &\tp     45.9 &        17.1
    &\tp     1.9 &\tp    19.2 &\tp     57.3 &\tp     45.6 &\tp     27.2
    &\tp     3.0 &\tp    27.3 &\tp     23.2 &\tp \bf 43.4 &        13.6
    \\
    \hline\hline
    Tarantula
    &     3.2 &    11.9 &     60.0 &     31.6 &     31.8
    &    96.0 &   573.6 &      2.5 &      1.5 &      2.5
    &     4.8 &    28.2 &     36.1 &     27.5 &     20.3
    &     2.4 &    17.2 &     49.6 &     29.9 &     24.6
    &     5.2 &    69.3 &     15.6 &      0.0 &     15.4
    \\
    Ochiai
    &     5.4 &    37.5 &     48.9 &     24.6 &     22.5
    &   108.1 &   922.6 &      1.3 &      0.5 &      1.8
    &     3.7 &    30.5 &     31.5 &     45.4 &     18.8
    &     2.5 &    19.8 &     46.8 &     55.8 &     28.7
    & \bf 2.5 &   254.7 &     19.7 & \bf 43.4 & \bf 19.8
    \\
    DStar
    &     5.7 &    86.2 &     38.5 &     22.9 &     20.6
    &   140.7 &   987.5 &      0.6 &      0.0 &      0.5
    &    10.3 &    29.4 &     28.9 &     48.1 &     17.3
    &     5.3 &    32.1 &     41.9 &     55.8 &     21.4
    &    12.7 &   402.1 &     23.7 & \bf 43.4 &     17.3
    \\
    Jaccard
    &     5.5 &    57.6 &     40.2 &     24.1 &     21.8
    &    98.9 &   586.3 &      0.8 &      0.0 &      1.4
    &     7.9 &    31.5 &     27.5 &     58.8 &     17.1
    &     3.6 &    18.2 &     47.9 &     49.7 &     23.5
    &     3.2 &   112.5 & \bf 26.3 & \bf 43.4 &     17.1
    \\
    GP13
    &    12.3 &   139.7 &     34.8 &      7.5 &     16.5
    &   445.4 &  1165.7 &      0.0 &      0.0 &      0.0
    &    12.6 &    30.7 &     28.2 &     46.5 &     15.6
    &    24.7 &    49.4 &     17.8 &     49.7 &      8.0
    &   413.4 &   659.5 &      0.0 &      0.0 &      0.0
    \\
    Naish2
    &    12.3 &   139.7 &     34.8 &      7.5 &     16.5
    &   445.4 &  1155.5 &      0.0 &      0.0 &      0.0
    &    12.6 &    30.7 &     28.2 &     46.5 &     15.6
    &    24.7 &    46.7 &     17.8 &     49.7 &      8.0
    &   413.4 &   659.5 &      0.0 &      0.0 &      0.0
    \\
    Overlap
    &     6.4 &    48.3 &     41.7 &     29.1 &     26.5
    &   430.9 &  1217.4 &      0.1 &      0.2 &      0.2
    &     6.3 &    40.6 &     26.0 &     24.0 &     17.0
    &    10.1 &    53.1 &     29.0 &     29.2 &     20.0
    &   262.3 &   576.7 &      0.0 &      0.0 &      0.0
    \\
    Harmonic
    &     5.7 &    30.4 &     55.3 &     34.4 &     30.7
    &   116.1 &   713.1 &      1.4 &      0.0 &      2.0
    &     3.8 &    29.1 &     31.8 &     35.3 &     19.2
    &     3.0 &    25.5 &     45.7 &     29.2 &     25.2
    &     4.9 &   224.7 &     18.3 &     34.7 &     14.5
    \\
    \hline
    Zoltar-s
    &     3.4 &    27.7 &     55.6 &     35.7 &     32.2
    &   264.6 &  1023.6 &      1.2 &      1.7 &      1.0
    &     3.6 &    29.4 &     31.7 &     35.3 &     19.3
    &     2.6 &    20.4 &     46.2 &     29.5 &     25.3
    &    79.9 &   439.4 &      4.2 &      0.0 &      5.2
    \\
    Hyperbolic
    &     8.1 &    42.4 &     45.6 &     24.1 &     19.5
    &   446.5 &  1147.9 &      0.0 &      0.0 &      0.0
    &     4.2 &    29.1 &     32.1 &     33.6 &     18.4
    &     8.2 &    40.2 &     27.9 &     23.4 &     17.9
    &   322.9 &   633.3 &      0.0 &      0.0 &      0.0
    \\
    Barinel
    &  1717.4 &  3288.4 &      0.1 &      0.0 &      0.0
    &  3772.9 &  4813.6 &      0.0 &      0.0 &      0.0
    &  1217.9 &  1915.2 &      0.1 &      0.0 &      0.0
    &  1863.5 &  3323.6 &      0.1 &      0.0 &      0.0
    &  1407.3 &  2557.0 &      0.0 &      0.0 &      0.0
    \\
    \hline\hline
    GRACE
    &    35.5 &    98.8 &      9.7 &     10.0 &      5.0
    & \multicolumn{5}{c||}{(aborted after 24 hours)}
    &     7.2 &    48.2 &     22.7 &     26.2 &     18.6
    &    22.2 &    87.9 &     12.5 &     10.5 &      5.0
    &   296.1 &   701.8 &      0.4 &      5.2 &      1.0
    \\\hline

  \end{tabular}
  }
  \end{table*}
  \begin{table*}
    \caption{
    Statement-level results for Defects4J multi-fault dataset
    averaged over all variants per project.
    See Table~\ref{tab:tcm-benchmarks} for notation.
  }
  \vspace*{-0.75ex}
  \centering
  {\footnotesize
   \setlength{\tabcolsep}{1.4pt}
   \setlength{\doublerulesep}{0.75pt}
  \begin{tabular}{|l||r|r|r|r|r||r|r|r|r|r||r|r|r|r|r||r|r|r|r|r||r|r|r|r|r|}\hline
    & \multicolumn{5}{c||}{Chart ($\mean{|S|}=67570$, $\mean{N}{\!_f}=3$)}
    & \multicolumn{5}{c||}{Closure ($\mean{|S|}=41969$, $\mean{N}{\!_f}=9$)}
    & \multicolumn{5}{c||}{Lang ($\mean{|S|}=9719$, $\mean{N}{\!_f}=6$)}
    & \multicolumn{5}{c||}{Math ($\mean{|S|}=23740$, $\mean{N}{\!_f}=5$)}
    & \multicolumn{5}{c|}{Time ($\mean{|S|}=10289$, $\mean{N}{\!_f}=7$)}\\
    \hline
    Metric
    &{\scriptsize AWE$_1$}&{\scriptsize AWE$_M$}&{\scriptsize R{\tiny @}10}&{\scriptsize P{\tiny @}1}&{\scriptsize P{\tiny @}5}
    &{\scriptsize AWE$_1$}&{\scriptsize AWE$_M$}&{\scriptsize R{\tiny @}10}&{\scriptsize P{\tiny @}1}&{\scriptsize P{\tiny @}5}
    &{\scriptsize AWE$_1$}&{\scriptsize AWE$_M$}&{\scriptsize R{\tiny @}10}&{\scriptsize P{\tiny @}1}&{\scriptsize P{\tiny @}5}
    &{\scriptsize AWE$_1$}&{\scriptsize AWE$_M$}&{\scriptsize R{\tiny @}10}&{\scriptsize P{\tiny @}1}&{\scriptsize P{\tiny @}5}
    &{\scriptsize AWE$_1$}&{\scriptsize AWE$_M$}&{\scriptsize R{\tiny @}10}&{\scriptsize P{\tiny @}1}&{\scriptsize P{\tiny @}5}  \\
    \hline
    FLITSR (T)
    &\tp     7.4 &\tp    60.1 &\tp     24.6 &\tp     24.0 &\tp     13.4
    &\tp    36.9 &     1806.6 &\tp \bf 15.5 &\tp \bf 38.8 &\tp \bf 21.7
    &\tp    35.5 &      151.5 &\tp      5.6 &\tp      2.4 &         1.7
    &\tp \bf 3.9 &\tp   130.5 &\tp \bf 34.3 &\tp \bf 36.1 &\tp\bf  16.6
    &\tp \bf 3.2 &      186.8 &\tp     18.5 &\tp \bf 56.5 &\tp     14.3
    \\
    FLITSR (O)
    &\tp \bf 7.1 &\tp   119.4 &\tp     25.3 &        26.2 &\tp     13.3
    &\tp    58.4 &     2920.3 &\tp     14.6 &         3.9 &\tp     15.5
    &\tp    28.8 &      175.4 &\tp      7.6 &         4.8 &\tp      3.5
    &\tp     7.0 &\tp   141.0 &\tp     24.8 &\tp     27.2 &\tp     11.4
    &\tp     5.7 &\tp   676.4 &\tp     18.8 &        21.7 &\tp \bf 15.6
    \\
    FLITSR (H)
    &\tp \bf 7.1 &\tp   106.7 &\tp \bf 25.7 &        26.2 &\tp \bf 14.4
    &\tp   126.2 &\tp  2561.1 &\tp     12.9 &\tp     17.8 &\tp     17.6
    &\tp    32.0 &      168.7 &\tp      7.5 &\tp      4.1 &\tp      3.6
    &\tp     4.2 &\tp    86.2 &\tp     31.2 &\tp     33.2 &\tp     15.2
    &\tp     6.1 &      582.0 &\tp     18.6 &\tp     43.4 &\tp     14.6
    \\
    \hline
    FLITSR* (T)
    &\tp     7.4 &\tp\bf 36.1 &\tp     24.6 &\tp     24.0 &\tp     14.2
    &\tp    22.2 &\tp\bf 1461.6&\tp \bf 15.5 &\tp \bf 38.8 &\tp \bf 21.7
    &\tp    31.0 &\tp   125.6 &\tp      5.6 &\tp      2.4 &         1.7
    &\tp \bf 3.9 &\tp   129.2 &\tp \bf 34.3 &\tp \bf 36.1 &\tp \bf 16.6
    &\tp \bf 3.2 &\tp    94.4 &\tp     18.5 &\tp \bf 56.5 &\tp     14.3
    \\
    FLITSR* (O)
    &\tp \bf 7.1 &\tp    39.9 &\tp     25.3 &        26.2 &\tp     13.3
    &\tp\bf 21.4 &\tp  1717.9 &\tp     14.6 &         3.9 &\tp     15.5
    &\tp\bf 26.0 &\tp\bf 110.7&\tp      7.5 &         4.8 &\tp      3.5
    &\tp     8.3 &\tp   142.3 &\tp     23.8 &\tp     27.2 &\tp     11.4
    &\tp     5.7 &\tp\bf 74.4 &\tp     18.8 &        21.7 &\tp \bf 15.6
    \\
    FLITSR* (H)
    &\tp \bf 7.1 &\tp    39.3 &\tp     25.6 &        26.2 &\tp     14.3
    &\tp    51.0 &\tp  2241.2 &\tp     13.0 &\tp     17.8 &\tp     17.6
    &\tp    30.2 &\tp   137.2 &\tp      7.5 &\tp      4.1 &\tp      3.6
    &\tp     4.2 &\tp\bf 84.4 &\tp     31.3 &\tp     33.2 &\tp     15.2
    &\tp     5.0 &\tp   136.0 &\tp \bf 20.6 &\tp     43.4 &\tp     14.6
    \\
    \hline\hline
    Tarantula
    &    20.3 &    64.6 &     16.2 &      6.3 &      6.6
    &    60.0 &  1803.9 &     13.2 &     13.7 &     12.5
    &    38.8 &   149.7 &      4.9 &      1.4 &      1.9
    &     7.2 &   145.1 &     26.1 &     14.2 &     12.9
    &     5.7 &   185.9 &     14.5 &      8.5 &     10.2
    \\
    Ochiai
    &    14.6 &   192.9 &     17.7 & \bf 26.9 &      6.0
    &   147.8 &  2916.2 &      2.4 &     13.7 &      2.4
    &    33.8 &   174.4 &      6.7 &      4.8 &      2.2
    &    10.1 &   162.9 &     19.4 &     17.6 &      7.8
    &    12.3 &   697.6 &     17.2 &     21.7 &     11.1
    \\
    DStar
    &    15.4 &   433.7 &     16.1 &     26.4 &      4.6
    &   528.5 &  3792.5 &      0.5 &      1.7 &      0.2
    &    76.1 &   193.6 &      8.1 &  \bf 6.0 &      2.1
    &    59.5 &   305.6 &     10.0 &     17.6 &      2.6
    &    82.6 &  1486.3 &     18.1 &     21.7 &     10.1
    \\
    Jaccard
    &    15.0 &   294.1 &     17.4 & \bf 26.9 &      5.1
    &   204.5 &  1865.2 &      0.6 &      1.7 &      0.6
    &    61.1 &   183.2 &  \bf 9.7 &  \bf 6.0 &      2.4
    &    13.6 &   163.8 &     15.3 &      2.1 &      4.3
    &    17.2 &   412.3 &     17.6 &     21.7 &     10.1
    \\
    GP13
    &    53.8 &   731.0 &     14.4 &     15.0 &      2.9
    &  2172.0 &  5869.1 &      0.3 &      0.5 &      0.1
    &    84.8 &   209.5 &      7.3 &      5.9 &  \bf 3.8
    &   251.6 &   449.6 &      2.9 &      1.2 &      0.8
    &  1477.4 &  2517.0 &      0.0 &      0.0 &      0.0
    \\
    Naish2
    &    53.8 &   731.0 &     14.4 &     15.0 &      2.9
    &  2172.0 &  5811.7 &      0.3 &      0.5 &      0.1
    &    84.8 &   209.5 &      7.3 &      5.9 &  \bf 3.8
    &   251.6 &   391.7 &      2.9 &      1.2 &      0.8
    &  1477.4 &  2517.0 &      0.0 &      0.0 &      0.0
    \\
    Overlap
    &    36.9 &   228.7 &     14.1 &      6.1 &      6.1
    &   628.7 &  5110.3 &      6.8 &      8.9 &      6.9
    &    50.3 &   258.7 &      2.9 &      1.5 &      1.5
    &    30.6 &   356.5 &     19.9 &     13.7 &     11.4
    &   240.2 &  2043.8 &      4.4 &      6.5 &      6.5
    \\
    Harmonic
    &    15.2 &   135.1 &     22.1 &     26.6 &     13.6
    &   126.9 &  2559.9 &     10.1 &     17.4 &      9.7
    &    33.6 &   167.2 &      3.5 &      2.6 &      1.7
    &     8.3 &   114.5 &     24.3 &     21.7 &     12.7
    &     6.3 &   581.7 &     15.8 &     21.7 &      9.0
    \\
    \hline
    Zoltar-s
    &    10.5 &   146.8 &     23.2 &     24.5 &     14.4
    &   336.6 &  3361.4 &     13.0 &     20.3 &     12.3
    &    32.7 &   171.2 &      3.6 &      2.6 &      1.7
    &     7.2 &   161.9 &     24.4 &     21.8 &     12.4
    &   129.1 &  1439.1 &      6.6 &      4.3 &      4.3
    \\
    Hyperbolic
    &    27.8 &   202.4 &     16.5 &     26.1 &      7.1
    &  1670.8 &  4480.0 &      0.3 &      0.5 &      0.1
    &    36.1 &   180.5 &      3.0 &      2.6 &      1.7
    &    54.6 &   294.7 &     13.1 &     19.4 &      9.5
    &  1052.8 &  2207.5 &      0.0 &      0.0 &      0.0
    \\
    Barinel
    &  9979.5 &  19104.6 &     0.0 &      0.0 &      0.0
    &  5859.1 &  25366.1 &     0.0 &      0.0 &      0.0
    &  4698.2 &   8210.1 &     0.0 &      0.0 &      0.0
    &  8109.5 &  18124.6 &     0.0 &      0.0 &      0.0
    &  1936.0 &   7493.5 &     0.0 &      0.0 &      0.0
    \\\hline
  \end{tabular}
  }
  \label{tab:d4j-results}
\end{table*}

\mysubsubsection{Synthetic Faults}
The results in \Cref{tab:tcm-benchmarks} show that all FLITSR and FLITSR*
variants perform very well for multi-fault versions ($N_f\geq
2$). AWE$_1$ ranges from 12-16 methods for $N_f=2$, depending
on the applied base metric, to approximately 3 methods (i.e., 0.5\% of the
average project size) for $N_f=32$, confirming
that it is easier to find \emph{a} fault if
there are more to choose from \cite{multipleInfluence}. $P@5$ grows similarly, from appproximately 18\% to
approximately 33\%, while $R@N_f$ remains mostly stable around 30\%.  However,
the AWE$_M$ values \emph{increase} with growing fault numbers, from approximately
30--40 for $N_f=4$ to approximately 45--65 for $N_f=32$, showing that the
multi-fault localization problem is hard indeed.

The highlighted entries show that for the multi-fault versions ($N_f\geq 2$)
\emph{all} FLITSR and FLITSR* variants substantially outperform all base
metrics in all measures, including Zoltar-s, Hyperbolic, and Barinel, which
were designed to handle multi-fault programs.  Moreover, the performance
difference generally grows with growing fault numbers. For example, for FLITSR*
using Harmonic as base metric,
the
AWE$_1$ improvement grows from 3.2\% (for $N_f=2$) to 42.5\% (for $N_f=32$),
AWE$_M$ from 29.7\% to 46.7\%, P@5 from 30.3\% to 60.2\%, and R@$N_f$ from
30.3\% to 60.2\%. However, in some cases the growth levels out at smaller fault
levels.
At $N_f=1$, FLITSR no longer outperforms all base metrics for all measures, but
any regressions remain small. This is expected because our approach targets
multi-fault systems.

\mysubsubsection{Defects4J}
\Cref{tab:d4j-method-results} shows that the results for the Defects4J dataset
vary considerably between the individual projects.
For Chart and Math, most base metrics
%
%
already perform well, but FLITSR still yields substantial AWE reductions to the
first (30\%-90\%) and median (30\%-75\%) faults, and similar increases in
recall (generally around 30\%) and precision (generally 20\%-50\%). AWE$_1$ is
below 2.0, with $P@1$ reaching up to almost 80\%, which means that
FLITSR pinpoints the top error in four out of five variants. FLITSR* yields
further improvements for AWE$_M$ of up to 70\% compared to the
base FLITSR, and 80\% compared to the corresponding base metric.
For the remaining three projects, FLITSR* shows drastic improvement over FLITSR,
i.e., these projects have dominating faults that require repeated localization
rounds.  For Lang and Closure FLITSR* yields improvements to the first
fault (e.g., a 50\% AWE$_1$ reduction for Closure), and reasonable improvements
to the median fault (10\%-30\%); for Time it yields better improvements
to the median fault (60\%-90\%), and no improvement for the first fault.

The bottom row of~\Cref{tab:d4j-method-results} shows the results for GRACE.
It performs much worse on our multi-fault Defects4J dataset variant than the
published results on the original Defects4J~\cite{grace} would indicate. It is
outperformed by all FLITSR-variants
on almost all measures; in particular, FLITSR and FLITSR* have much
smaller AWEs to the first and median faults. This may indicate that GRACE may not
generalize well to multiple fauts.

\label{sec:eval:d4j:results}

\subsection{Statement-Level Localization Results}
\label{sec:eval:d4j}
%

\begin{figure}
  \hspace*{-1.7ex}
  \includegraphics[scale=0.37]{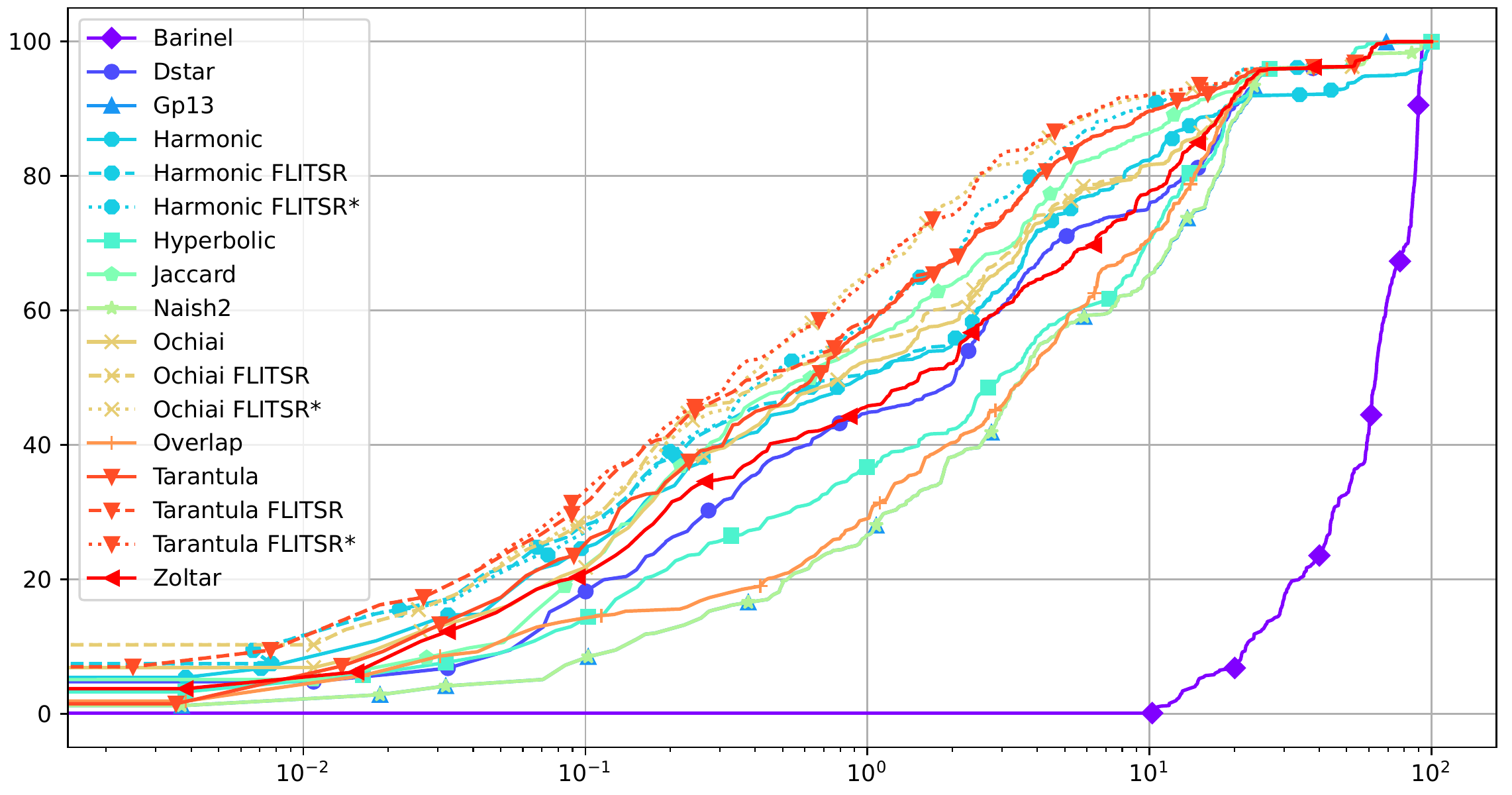}
  \caption{
    Percentage of bugs found over proportion of statements inspected in
    rank order, averaged over Defects4J multi-fault dataset.
    x-axis shown in log-scale.
  }
  \label{fig:statement-summary}
\end{figure}

\Cref{fig:statement-summary} shows the statement-level results for the
Defects4J dataset. Overall, it confirms the method-level results (cf.\
\Cref{fig:tcm-summary}): FLITSR* outperforms FLITSR for each base metric, which
in turn outperforms the base metric itself. Since there are generally 10$\times$
more statements than methods, the AWEs are still higher and precision and recall
values are lower (cf.\ \Cref{tab:d4j-results}).

The individual project results in \Cref{tab:d4j-results} show that the FLITSR*
variants indeed consistently outperform the base metrics.

\subsection{Statistical Analysis}
%
We performed a non-parametric paired significance test (Wilcoxon signed-rank test)
on the AWE numbers. For the synthetic faults, this shows that both FLITSR-variants
statistically significantly (p<<0.05) outperform the respective base metrics in
aggregate over all fault densities (and FLITSR* outperforms FLITSR); however, for
single-fault programs and lower fault densities ($N_{\!f}=2$) the differences for some
individual metrics are not statistically significant. For real faults,
both FLITSR-variants again statistically significantly outperform the respective
base metrics to the first fault (when aggregated over all projects), and while
FLITSR* statistically significantly outperforms the base metrics to the median
fault, FLITSR does not.

\subsection{Threats to Validity}
%
\label{sec:eval:ttv}
%

\mysubsubsection{Internal Validity}
\label{sec:ttv:internal}
%
There are several threats to the internal validity of our
evaluation.
We carefully tested and documented our implementation, data collection
and evaluation scripts to mitigate against \emph{implementation errors}. We
also make a {replication package}~\cite{replication} available. We ran GRACE as
a black-box system using its default settings, which may not be
optimal, but due to the complexity of its implementation we cannot tell.

The main threat is \emph{ground truth} errors in the execution spectra and the
actual fault locations.
For the method-level dataset, we cannot judge whether the TCMs
faithfully reflect actual system executions, and we have to accept them as
ground truth.
%
For the Defects4J dataset, we relied on GZoltar \cite{gzoltar}
to collect execution spectra, which is widely used in the literature
(e.g., \cite{DBLP:conf/icse/PearsonCJFAEPK17, DBLP:conf/sigsoft/XuanM14,
DBLP:conf/wcre/LiuK0B19}), but is prone to potential imprecisions in the
collection of statement-level execution data, which may affect the results.
We discard non-executed statements labelled as fault locations, and discard faults
that only affect non-executed statements to minimize bias effects.
Similarly, our fault location oracle is based on the statements that are
identified by the translation process (cf.\ \Cref{sec:eval:d4j:setup});


Finally, \emph{residual faults} not exposed by the test suites can still impact
execution through fault masking.
However, any such {residual faults} affect all applied approaches (if not
necessarily to the same extent), which mitigates
their impact.

\mysubsubsection{External Validity}
\label{sec:ttv:external}
%
Our results may not necessarily generalize beyond our experimental set-up.
The main threats stem from the
\emph{execution spectra}, the \emph{fault types}, and the \emph{applied
metrics}.
%
Our evaluation
uses simple program execution spectra derived from Java-based systems.
Spectra derived from other languages \cite{dstar,DBLP:journals/jss/AbreuZGG09,
DBLP:conf/issre/WangSMK20,DBLP:conf/sigsoft/KhanSW21,
DBLP:journals/tosem/TroyaSPC18,DBLP:conf/sle/RaselimoF19} or
involving more complex types of elements (e.g., data flow edges
\cite{DBLP:conf/icse/SantelicesJYH09}) could have different
characteristics that impact the results.

Our method-level evaluation partly relies on
mutation-based fault injection \cite{threats}.
There is some evidence
\cite{DBLP:conf/sigsoft/JustJIEHF14,DBLP:conf/icse/PapadakisSYB18} that they
may be a valid substitute for real faults in software testing, but
\Cref{fig:tcm-summary} shows that the localization results differ
substantially, although the general trends remain similar.
Tables \ref{tab:d4j-method-results} and \ref{tab:d4j-results} show that
different real-world projects lead to quite different results; other projects
with different fault characteristics
may not support our conclusions.

Since our results vary with the base metric, using different metrics could lead
to different results. However, we chose metrics that are widely used in the
literature (cf.\ \cite{DBLP:conf/icse/PearsonCJFAEPK17}) and have different
characteristics. Likewise, other learning-based systems may produce better
results than GRACE. However, we were unable to get any other system to run over
our datasets, and GRACE was the best performing system investigated in
\cite{grace}.

%
%

\subsection{Answers to Research Questions}
%
\label{sec:eval:summary}
\label{sec:eval:steimann:summary}
\label{sec:eval:d4j:summary}
We can now summarize the results of our experimental evaluation into answers to
our research questions.

\textbf{RQ1} (comparison with base metrics):
While the specific results vary with the dataset, measurement, and base metric,
FLITSR generally substantially improves over the base metric, and regresses
only in a few cases and only by small amounts. AWE improvements reach 90\%,
recall improvements 80\%, and precision improvements 60\%; in most
cases we see improvements between 20\% and 50\%

\textbf{RQ2} (effects of choice of base metric):
For the synthetic fault dataset, we see in \Cref{fig:tcm-summary} the respective
curves for FLITSR and in particular FLITSR* using different base metrics run very
closely together, even when the base metrics themselves differ drastically (e.g.,
Ochiai and Tarantula). We can therefore conclude that FLITSR normalizes the variations
in the base metrics, and even a bad base metric will give a good FLITSR (and FLITSR*)
result. Nevertheless, the best results are achieved using Harmonic and
Ochiai, which are both well-performing single-fault metrics \cite{sbflSurvey}.

For the much smaller real fault dataset we see bigger differences between the
metrics, both at the method- and statement-level; in particular, Tarantula now
emerges as strong base metric as well.

\textbf{RQ3} (comparison with learning-based localization):
On our multi-fault Defects4J dataset GRACE performed much worse on all measures
than FLITSR using almost any base metric. This demonstrates that FLITSR
outperforms even the state-of-the-art learning-based techniques.

\textbf{RQ4} (comparison between FLITSR and FLITSR*):
For both real and injected fault datasets, it is clear that FLITSR* improves upon
FLITSR when there are more dominated or masked faults, which is the case in most
versions. For the synthetic fault dataset, the improvements of FLITSR* over FLITSR
are more consistent (with all fault levels except $N_f=1$ showing improvements),
but are less pronounced.
For the real fault dataset we see larger
improvements that are very project dependent, which indicates which projects
have higher fault masking occurring.

\textbf{RQ5} (effects of fault type and coverage granularity):
%
\Cref{fig:tcm-summary} shows that
the synthetic fault dataset provides more normalized, consistent results than real faults, with
FLITSR and FLITSR* still outperforming all other metrics, and the FLITSR*
variants that perform the best remaining roughly the same. However, this could
be attributed to the large number of versions
(75000) normalizing the results.
%
A comparison of the method- and statement-level results
in \Cref{tab:d4j-method-results} and \Cref{tab:d4j-results} shows that AWEs to
all faults increase, and the
$P@X$ generally decrease for the statement level, which is expected
as there are more elements to consider.

%% file: relwork.tex
\section{Related work}
\label{sec:relwork}

 \mysubsubsection{Spectrum-based fault localization}
%
%
The first widely used SBFL metric was Tarantula \cite{tarantula}; following its
introduction, a large number of different metrics (in total, 35 
metrics are listed in the surveys by Wong et al.~\cite{survey} and de~Souza et
al.~\cite{sbflSurvey}) with different characteristics have been proposed and
evaluated. While there is no consensus about a ``best'' metric, Tarantula,
Ochiai \cite{ochiai}, Jaccard \cite{jaccard}, and DStar \cite{dstar}
are widely used in current work.
%
Zhang et al.\ investigate using the PageRank algorithm to further improve the SBFL
rankings~\cite{pagerank}.

\outdated{
====== OLD ======

There have been many SBFL ranking metrics proposed stemming from a variety of
fields. The first to be developed and most widely recognised being Tarantula, proposed
by Jones et al.\ in 2002 \cite{tarantula}. The Tarantula metric was created for
use in a visualization tool, developed to assist developers in the debugging process.
Other notable mentions include Ochiai \cite{ochiai}, a metric taken from the molecular
biology domain and adapted by Abreu et al.\ for the fault localization setting; DStar,
proposed by Wong et al.\  and developed from a binary similarity coefficient-based
analysis standpoint; and Jaccard \cite{jaccard}, developed as a data clustering coefficient
and implemented in Pinpoint, a fault detection framework for root cause analysis.
}

 \mysubsubsection{Multiple-fault localization}
%
Over the last decade, much of the focus of fault localization has shifted to
multi-fault localization, where three common approaches have
emerged: \emph{one-bug-at-a-time} (OBA), \emph{parallel debugging}, and
\emph{multiple-bug-at-a-time} (MBA)~\cite{DBLP:journals/infsof/ZakariLAAR20}
(see \Cref{sec:intro}).
The most popular is \emph{one-bug-at-a-time} (OBA)
\cite{DBLP:conf/wcre/LiuLNBB16,DBLP:conf/icst/SunP16,DBLP:conf/icst/PerezAd17}
for which our evaluation
shows that even the AWE to the first fault is improved by FLITSR,
which would inevitably aid OBA when it is applied.
The second approach is \emph{parallel debugging}.
Various techniques in this domain have been
investigated~\cite{parallel:jones, parallel:podgurski, parallel:steimann,
DBLP:journals/smr/ZakariL19}, and many report promising results, however
many issues arise due to this method, such as Bug Races and lost test cases
and elements. In addition to this, the parallel debugging technique assumes a
truly parallel setup can be achieved in practice, and thus addresses a different
problem.
\outdated{
TODO, do we want the following:
In the light of this
research, our work can be described as creating one such paralellization that uses
the SBFL metric employed to group related elements and tests, similar to that of
the fault localization clustering algorithm found in \cite{parallel:jones}.
}
The last type of multiple-fault localization is multiple-but-at-a-time (MBA),
under which our technique falls.
The research done for MBA mostly employs additional technologies to
aid the SBFL results, such as logic reasoning approaches that originate from
MBFL~\cite{zoltarM,DBLP:journals/jss/AbreuZG11} (which we compare to), genetic
algorithms~\cite{DBLP:journals/jss/ZhengWFCY18}, a linear programming
model~\cite{linBasis}, and ideas from Complex Network
Theory~\cite{DBLP:journals/access/ZakariLC18}. All of these use technologies
with large time complexities making them somewhat infeasible in practice.
Lastly, the approach by Steimann and Bertschler~\cite{numFaults} requires the probability
distribution of the number of faults as an additional input. Our approach uses
no extra information or additional technologies, providing a simple but powerful
solution.

 \mysubsubsection{Learning-based SBFL}
%
In addition to traditional SBFL, two classes of learning-based SBFL techniques
have been proposed,
\begin{enumerate*}
  \item learning-to-represent, which learn relationships in finer coverage
    spectra~\cite{DBLP:journals/tr/WongDGXT12,
    DBLP:journals/ijseke/WongQ09,DBLP:conf/wcre/ZhangLML19}, and
  \item learning-to-combine, which learn to combine multiple fault localization
    techniques~\cite{deepFl,DBLP:journals/pacmpl/LiZ17,DBLP:conf/issta/SohnY17}
\end{enumerate*}.
We experimentally compare FLITSR to GRACE~\cite{grace} (a combination of both techniques) as the
state of the art for the learning based techniques.

\mysubsubsection{Test-suite reduction}
As test suite sizes increase during software evolution, so does the cost of test
suite execution. Different test suite reduction techniques~\cite{testSuiteReduction1,
testSuiteReduction2, testSuiteReduction3} have been developed to counter this.
However, these techniques focus on improving the test suite execution efficiency, while
FLITSR focuses on improving fault localization results.

%% file: conclude.tex
\section{Conclusions and future work}
\label{sec:conc}
\subsubsection*{Conclusions}
We present a novel MBA approach called FLITSR (and its multi-iteration variant FLITSR*)
to address the problems arising when popular SBFL techniques are applied to
multi-fault programs. In particular, since
the bug prioritization order
in practice differs from
the ranked list of faulty elements, it may be necessary to consider
faults other than the top-ranked fault. This highlights the need for a set of
high-priority elements such as the basis returned by FLITSR, and the importance
of improving the localization of \emph{multiple faults} in the ranking.
%
We applied our approach to many underlying base metrics
developed for both single- and multi-fault scenarios, and evaluated this on two
datasets, one at the method level containing injected faults, and one at the
method and statement levels containing
real-world faults.

For the synthetic method level fault dataset, we found that FLITSR
greatly reduces the AWE to the first fault (up to 60\%),
especially at a higher number of faults. We also saw that FLITSR* further
reduces the AWE to the median fault, by about half on average, and almost doubles precision and recall.
%
%
On the Defects4J real-fault dataset, we found that
the improvements of FLITSR are
more pronounced (up to 90\% and 75\% reduction in AWE$_1$ and AWE$_M$ resp.),
but the variation in the performance of FLITSR
using differing base metrics also increases, as does the variation in
performance of the base metrics themselves. The choice of base metric thus appears to be
more important for real-world data, as a bad choice can lead
to suboptimal results. The wasted effort is significantly
higher for the statement level than for the method level, which is expected as there
are many more elements to consider. However despite
this, there is little variation in the improvements achieved by FLITSR between the
statement and method level on the Defects4J dataset. This indicates that
FLITSR is general in the level of coverage used, and can provide improvements in
both cases.

The comparison of FLITSR to the state-of-the-art learning-based tool GRACE shows
that FLITSR not only outperforms the SBFL metrics it uses, but also other, more
recent techniques. This indicates that it extends the localization induced
by any metric to the multi-fault case, so that it outperforms state-of-the-art
techniques, even if the base metric does not originally perform well for multiple
faults.

 \mysubsubsection{Future work}
The tie resolution problem present in existing SBFL techniques \cite{ties1, ties2}
is
amplified in FLITSR.
We have used the original SBFL metric's
rankings as a tie-breaker for the evaluation here. We will
consider better tie-break strategies in future work.

%
Although
our evaluation shows that FLITSR improves
considerably upon the base metrics in the multi-fault setting,
we do not know what effect this will have on developers in practice. We also do not know
whether the classic ranked list or a minimal basis output will be more
useful.
We therefore plan to deploy FLITSR in a real-world setting.
%

Lastly, many other fault localization techniques 
also produce rankings similar to those of the base metrics
and can thus be applied in conjunction with FLITSR
to potentially realize further improvements, and we plan to further
evaluate this.
